\documentclass[12pt]{article}
\usepackage{amsmath,a4wide}
\usepackage{graphics,graphicx}
\usepackage{hyperref}
\usepackage{cite}
\usepackage{epsfig,psfrag}

\newcommand{\la}{\langle}
\newcommand{\ra}{\rangle}
\newcommand{\tr}{\mathrm{tr}}
\newcommand \vev [1] {\langle{#1}\rangle}
\newcommand{\cL}{{\cal L}}
\newcommand{\cO}{{\cal O}}
\newcommand{\nt}{\notag\\}
\newcommand{\cN}{{\cal N}}
\newcommand{\p}[1]{(\ref{#1})}
\newcommand{\q}{\theta}
\newcommand{\cT}{{\cal T}}
\newcommand{\cA}{{\cal A}}

\newcommand \bra [1] {\langle {#1}|}
\newcommand \veve [1] {\langle{#1}\rangle}
\newcommand{\bq}{\bar\theta}
\newcommand{\da}{{\dot\alpha}}
\renewcommand{\a}{{\alpha}}

\csname @addtoreset\endcsname{equation}{section}
\begin{document}

\thispagestyle{empty}

\null\vskip-43pt \hfill
\begin{minipage}[t]{50mm}
CERN-PH-TH/2011-061 \\
DCPT-11/11 \\
IPhT--T11/037 \\
\end{minipage}

\vskip1.5truecm
\begin{center}
\vskip 0.2truecm 

 {\Large\bf
The super-correlator/super-amplitude duality: Part II}
\vskip 1truecm

{\bf    Burkhard Eden$^{a}$, Paul Heslop$^{a}$, Gregory P. Korchemsky$^{b}$, Emery Sokatchev$^{c,d,e}$ \\
}

\vskip 0.4truecm
$^{a}$ {\it  Mathematics department, Durham University, 
Science Laboratories,
 \\
South Rd, Durham DH1 3LE,
United Kingdom \\
 \vskip .2truecm
$^{b}$ Institut de Physique Th\'eorique\,\footnote{Unit\'e de Recherche Associ\'ee au CNRS URA 2306},
CEA Saclay, \\
91191 Gif-sur-Yvette Cedex, France\\
\vskip .2truecm $^{c}$ Physics Department, Theory Unit, CERN,\\ CH -1211, Geneva 23, Switzerland \\
\vskip .2truecm $^{d}$ Institut Universitaire de France, \\103, bd Saint-Michel
F-75005 Paris, France \\
\vskip .2truecm $^{e}$ LAPTH\,\footnote[2]{Laboratoire d'Annecy-le-Vieux de Physique Th\'{e}orique, UMR 5108},   Universit\'{e} de Savoie, CNRS, \\
B.P. 110,  F-74941 Annecy-le-Vieux, France
                       } \\
\end{center}

\vskip -.2truecm 

\centerline{\bf Abstract}
\medskip
\noindent 
We continue the study of the duality between super-correlators and scattering super-amplitudes in planar $\cN=4$ SYM. We provide a number of further examples supporting the conjectured duality relation between these two seemingly different
objects.  We consider the five- and six-point one-loop NMHV and the six-point tree-level NNMHV amplitudes, obtaining them from the appropriate correlators
of strength tensor multiplets in $\cN=4$ SYM. In particular, we find
exact agreement between the rather non-trivial parity-odd sector of
the integrand of the six-point one-loop NMHV amplitude, as obtained
from the correlator or from BCFW recursion relations. Together these
results lead to the conjecture that the integrands of any N${}^k$MHV
amplitude at any loop order in planar $\cN=4$ SYM can be described by the correlators of stress-tensor multiplets. 

\newpage

\section{Introduction}

In the maximally supersymmetric Yang-Mills theory in four dimensions
($\cN = 4$ SYM) there is a duality between scattering amplitudes and
Wilson loops with light-like edges. This was first noticed at strong
coupling \cite{aldMald} via the AdS/CFT correspondence \cite{123}, and
soon after confirmed also at weak coupling \cite{dual11,dual12,dual21,dual22}
directly within the field theory.

Recently, it has been realised that both objects, Wilson loops and MHV
amplitudes, can be obtained from the { light-cone limit of}
correlation functions of certain gauge invariant {scalar} composite operators, { which are the bottom component of the $\cN=4$ stress-tensor multiplet $\cT$} \cite{AEMKS,EKS2}.

In other recent developments, a procedure for computing the  \emph{integrand}
of all scattering amplitudes in the theory (i.e. for 
all helicities and at all loop orders) has been derived \cite{nima1}
in terms of momentum 
twistor variables~\cite{hodges}  using 
 BCFW recursion relations~\cite{BCFW} generalized to loop
 level~\cite{Boels:2010nw,nima1}. Supersymmetric 
generalisations of the polygonal Wilson loops have also been suggested
as a dual to non-MHV amplitudes in two publications \cite{marksAndSparks,
simon} (see also recent comments in \cite{Belitsky:2011zm}).

For MHV amplitudes the equivalence to correlation functions of scalar
operators holds for the
integrands, which was verified in \cite{EKS3}. In this and the twin paper
\cite{twin} we propose to extend this duality to all non-MHV super-amplitudes
 and to the  light-cone limit of the super-correlators of stress-tensor multiplets,
respectively. We argue that the integrands
of all planar amplitudes are contained as a subsector of the correlation functions.
As an illustration of our proposal, in~\cite{twin} we demonstrated that the $n$-point
tree-level NMHV super-amplitudes can be obtained from the aforementioned
correlators, computed at tree level. Here we continue our investigation of the
 new duality. We set out to show the same for the NMHV five- and six-point
amplitudes at one loop, and for the NNMHV six-point amplitudes at tree-level.

The conjectured duality can be formulated as follows. Consider  the
 super-correlation
functions of $n$ energy momentum supermultiplets $\la \cT(1) \ldots
\cT(n)\ra$ in $\cal N$=4 superspace
in the limit in which consecutive points become light-like separated. This
 correlator depends both on the chiral ($\q$) and anti-chiral ($\bq$) odd
 coordinates of $\cN=4$ superspace. To be able to compare it to the
 super-amplitudes $\cA_n$ defined in {\it chiral} dual superspace $(x,\q)$,
 we set all $\bq=0$. Further, before taking the light-cone limit, we divide
 the correlator by its bottom component 
$\la \cT(1) \ldots \cT(n) \ra^{\mathrm{tree}}_{n;0}$, obtained by setting
 $\q=\bq=0$ and computed at tree level. This removes the pole singularities
 due to propagator factors.
  Then we claim that the light-cone limit of the ratio of
correlators is equivalent to the {\it square} of the  planar super-amplitudes
$\cA_n$ divided by the tree-level MHV amplitude\footnote{To obtain a
well-defined ratio the delta functions {imposing the super-momentum conservation} are removed from the
amplitudes.}
\begin{align}
\lim_{x^2_{i \, i+1} \rightarrow \, 0} \, \frac{\la
    \cT(1) \ldots \cT(n)\ra} {\la \cT(1) \ldots \cT(n)
    \ra^{\mathrm{tree}}_{n;0}} \bigg|_{\bar \theta_i=0}
=
\Big({ \cA_n}/ { \cA^{\text{tree}}_{n ;{\text{MHV}}}} 
\Big)^2\,. \label{corConj1}
\end{align}
At the moment this is slightly schematic and much of this equation
needs to be defined more carefully in order for the 
reader to be able to properly interpret it (e.g. on what (super)space
are the two sides defined and how are they related etc.) and we will do this
carefully in the next
section.  Note now however one intriguing feature. The left-hand side
is {\em not equal 
to} the correlation function directly but the correlation function
with a coupling dependent  rescaling of the odd coordinates $\theta \rightarrow a^{-1/4} \theta $,  with $a= g^2 N_c/ \pi^2 $ being the 't Hooft coupling. A similar rescaling of odd coordinates was performed in~\cite{simon} to compare the  supersymmetric Wilson loop with the superamplitude.

The other thing to note now is that { the quantities on both sides of the duality \p{corConj1}} diverge at
loop level and need regularising. However, on the correlation function
side loop corrections can be computed by considering integrals of
tree-level correlators with  multiple insertions  of the $\cN=4$ SYM Lagrangians
(itself a member of the stress-tensor multiplet). 
This enables us to define the left-hand side of the duality at the level of the
integrand via a tree-level {\it rational} correlation function. It can
then be compared with the  {\it rational} integrand for the entire super-amplitude, and
we find complete agreement for every test performed so far.

Together these results lead to the conjecture that the integrands of
any N${}^k$MHV amplitude at any loop order in planar $\cN=4$ SYM can
be described by the correlators of stress-tensor multiplets.

The paper is organised as follows. In Section \ref{sec:duality} we give a summary of the formulation of the new duality (for more detail see \cite{twin}). In Section~\ref{useESS} we exploit the  off-shell one- and two-loop $4$-point
correlator results of \cite{ESS}  (lifted to $\cN = 4$ as in
\cite{me1,partialNonRen}) 
to obtain the five-point tree NMHV, the five-point one-loop NMHV and
the six-point tree NNMHV 
amplitudes from our conjecture (\ref{corConj1}). In Section~\ref{newIntCheck} we construct the
six-point one-loop NMHV integrand from (\ref{corConj1}), which is much more
involved than the five-point case because there is a large parity-odd sector.
Using the same techniques as in \cite{EKS3} we verify exact agreement with
the result of \cite{nima1,nima2} based on BCFW recursion
relations.

\section{The duality}
\label{sec:duality}

The correlation functions in the new duality naturally depend on  chiral 
and anti-chiral Grassmann odd variables, while the amplitudes are usually
formulated on chiral superspaces \cite{Nair,annecySuperspace}. We argue in
 \cite{twin} and here that the amplitudes are found in the purely
left-handed sector of the correlators. In the present paper we focus
on explicit calculations; the interested reader can find a more complete
exposition of the various superspaces and superfields in  \cite{twin}.

The field content of the ${\cal N} = 4$ super Yang-Mills theory comprises
six real scalars, four complex Majorana-Weyl fermions and the gauge potential
$A_\mu$. The associated field strength, the scalars and the fermions all
transform in the adjoint representation of the gauge group, which we assume
to be $SU(N_c)$. A particularly useful way of presenting the multiplet on
shell is via $\cN = 4$ \emph{analytic superspace} \cite{paulN4}. In
this formalism the entire multiplet  can be sandwiched into a single scalar
superfield, { charged under a $U(1)$ subgroup of $SU(4)$}
\begin{equation}
W_{{\cal N} = 4}(z) \, , \qquad z \, = \, \{ {x^{\dot \alpha\a}}, \, {\rho^{\alpha  a} },
\, \bar \rho_{\, a'}^{\ \; \dot \alpha}, \, {y_{a'}}^a \} \, , \qquad a \in \{1,2\}
\, , \quad a' \in \{3,4\} \, .  \label{defAna}
\end{equation}
Here $\rho$ and $\bar \rho$ are odd variables~\footnote{The variables
  $\rho, \bar \rho$ each have just 4 components. This is half the
  number one would expect for $\cal N$=4 supersymmetry. This is thus
 similar to chiral superspace in which (the anti-chiral)   half of the odd coordinates
 are dropped.} 
 and $y$ is an additional bosonic 
coordinate related to the internal $SU(4)$ symmetry group. The $\rho$
variables are harmonic projections of the full Minkwoski superspace
variables $\theta$. For instance 
$\rho_i^{\alpha a} := \q_i^{\alpha a} + \q_i^{\alpha a'} y_{i\,a'}{}^a$
(more information can be found in appendix~\ref{sec:nairs-eta-as}).\footnote{ In \cite{twin} we used the alternative, harmonic superspace notation. There the variables $y$ are part of an $SU(4)$ harmonic matrix $u^{+a}_A, u^{-a'}_A$, and $\rho^a_\a$ is equivalent to $\q^{+a}_\a = \q^A_\a u^{+a}_A$.}

$\cal N$=4 analytic
superspace is the most convenient formalism for packaging together correlation functions
in $\cal N$=4 SYM since it manifests the full superconformal symmetry of
the problem enabling one to completely solve the superconformal Ward
identities and write any correlation function in  a fully
superconformal way~\cite{Heslop:2001dr}. On the other hand, $\cN=4$ SYM does not
have an off-shell { superspace} description and so in order to perform actual perturbative
calculations one needs to use $\cal N$=2 harmonic
superspace~\cite{GIKOS} and then lift the results to $\cal N$=4 analytic superspace.

The stress-tensor multiplet contains, among others, the following components in its $\rho,\bar\rho$ expansion: 
\begin{equation}\label{TT}
\cT(x,\rho, \bar\rho, y) \, = \, \tr( W_{{\cal N} = 4}^2 ) =  \cO + \ldots - 4  \rho^4 \cL  +
\ldots  - 4  \bar\rho^4 \bar\cL + \ldots +
(\rho\sigma^\mu\bar\rho)(\rho\sigma^\nu\bar\rho)T_{\mu\nu} + \ldots \, ,
\end{equation}
where 
\begin{equation}
\rho^4 \, = \, \frac{1}{12} \, (\rho^2)^{\alpha \beta} \,
(\rho^2)_{\alpha \beta} \, , \qquad (\rho^2)^{\alpha \beta} \, = \,
\rho^{a \alpha} \rho_a^\beta \label{defRho2and4}
\end{equation}
and so $\rho^4 \, = \, -   (\theta^+)^4/12$ in the notation of \cite{twin}. 
Here the lowest component $\cO=\tr(\phi^2) = \cT(x,0,0,y)$ is a scalar bilinear operator in the $\mathbf{20'}$ of $SU(4)$. The top spin component $T_{\mu\nu}$ is the stress tensor of the theory, which gives the multiplet its name. Another component of $\cT$, of crucial importance in what follows,  is the
chiral on-shell $\cN=4$ SYM Lagrangian $\cL$ appearing at $\rho^4$ (as well as its PCT conjugate $\bar\cL$ at $\bar\rho^4 $). For ease in later formulae, we { absorb the nilpotent factor into the definition of $\cL$:}
\begin{equation}\label{eq:12}
\rho^4 \cL \, \to  \,  {\cal L} \, .
\end{equation}
Due to a residual $\mathcal{Z}_4$ $R$-symmetry of the theory { (the centre of $SU(4)$)},
the expansion of $n$-point functions of the stress-tensor
multiplet $\cT$ in terms of the Grassmann variables is organised in powers
$\rho^{m} \bar \rho^{n}$ with $m-n=4k $ divisible by four \cite{EHW}.

In the present article we will not be interested in the right-handed spinors
$\bar \rho$ which we put to zero. The right-handed Poincar\'e supersymmetry
$\bar Q$ and the left-handed conformal supersymmetry $S$ of the model
are explicitly broken by this choice\footnote{However, since for the dual
amplitudes in formula (\ref{corConj1}), the full { dual superconformal
symmetry~\cite{annecySuperspace}} is present at tree
level~\cite{Brandhuber:2008pf}, then for the correlation functions
also in the Born approximation and in the light-like $n$-gon limit these 
symmetries  should be ``magically'' restored.}.

Then the entire correlator at $\bar \rho = 0$ is expanded in terms of
polynomials in $\rho$, homogeneous of degree $4 \, k$:
\begin{equation}
\la \cT(1) \ldots \cT(n) \ra|_{\bar\rho_i=0} \, = \, \sum_{k=0}^{n-4} G_{n;
  k}(1,\ldots,n;a)\ .
\label{corLeft}
\end{equation}
In what follows we will not display the restriction $\bar\rho_i=0$ explicitly,
but it will always be assumed. 
We can use the left-handed  Poincar\'e supersymmetry $Q$ and the
right-handed conformal 
supersymmetry $\bar S$ to simultaneously put $\rho_i = 0$ at any
four  points.  This explains the range $0 \leq k \leq n-4$. For example, at five points we could put $\rho_1= \ldots=\rho_4 = 0$
leaving only $\rho_5$, so that
the only possible terms in the expansion have $(\rho_5)^0$ and $(\rho_5)^4$ times
some functions of the bosonic coordinates $x^{\da\a}$ and $y_{a'}^a$.  The dependence on the full
set of $\rho_i$ can eventually be reconstructed by the inverse supersymmetry
transformation.

We also have an expansion in the 't Hooft coupling $a = g^2 N_c / \pi^2$
 and so we write the full $n$-point correlator as the double expansion
 \begin{equation}
\la \cT(1) \ldots \cT(n) \ra \, = \, \sum_{k=0}^{n-4} \sum_{l=0}^\infty a^{l+k}
  \, G^{(l)}_{n;k}(1,\ldots,n)\ ,
\label{corLeftloop}
\end{equation}
so that we denote by $G_{n;k}^{(l)}$ the $n$-point correlator
at Grassmann level  $O(\rho^{4k})$ and at $l$ loops.

The lowest contribution to $G_{n;k}$ -- so $G_{n;k}^{(0)}$, which we shall 
call the Born level -- comes at $O(a^k)$ from $(k+1)$-loop graphs
w.r.t. ordinary momentum space loop counting. The $(l)$ counter labels 
the order beyond Born approximation. In the correlation functions
$G_{n;k}^{(l)}$ thus carries $a^{(l+k)}$, quite different from
the corresponding amplitude as we discuss shortly. 

However, it is natural to gather together all the $(l)$ contributions to
the correlator (even though they occur at different powers of the coupling).
So for example we will define
\begin{equation}
  G_n^{(l)}(1,\dots,n):=
\  \sum_{k=0}^{n-4}  \, G^{(l)}_{n; 
  k}(1,\ldots,n)\ 
\end{equation}
to be simply the sum of all the $(l)$ contributions to the
$n$-point correlator. 

Now we compare the expansion of the correlator with the total colour ordered
$n$-point planar scattering amplitude $\cA_n$ (i.e. the sum of the MHV, NMHV,
... parts). This has an expansion very similar to the correlator 
(\ref{corLeft}):
\begin{equation}\label{eq:2}
\frac{\cA_n}{\cA_{n \, \text{MHV}}^{\text{tree}}} \,
\, = \, \sum_{k=0}^{n-4} \widehat \cA_{n; k} \ , 
\end{equation}
where the ratio is understood in the sense of removing the momentum and supercharge conservation delta functions. The amplitude is a function of three equivalent sets of variables. These can either be $\lambda_i^\a$, $\tilde\lambda_i^\da$ and $\eta_i^A$  (with $A=1,2,3,4$) of the chiral on-shell superspace \cite{Nair}, or $x_i^{\da\a}, \q^A_{i\,\a}$ of the chiral dual superspace \cite{annecySuperspace}, or $\lambda_i^\a, \mu_{i\,\da}, \chi_i^A$ of momentum supertwistor space ~\cite{Mason:2009qx}. The
bosonic variables $x$ are ``$T$-dual'' to the outgoing on-shell particle momenta \cite{xCoords,aldMald}:
\begin{equation}
(p_{i})^\alpha_{\dot \alpha} \, = \, \lambda^\alpha_i \bar \lambda_{i \, \dot \alpha}
\, = \, (x_{i}  - x_{i+1})^\alpha_{\dot \alpha} \, = \, 
(x_{i \, i+1})^\alpha_{\dot \alpha}\ .
\end{equation}
For the purpose of comparing with super-correlators, it is most convenient to use the momentum supertwistor odd variable $\chi^A =\lambda^\a \q^A_\a$. It is a Lorentz
scalar but it carries a four-component internal index $A = (a,a')$. Hence it has
the same number of  odd components as $\rho^\alpha_a$. 

The loop expansion is more straightforward than for the correlator, we have the double expansion 
\begin{equation}\label{eq:2}
\frac{\cA_n}{\cA_{n \, \text{MHV}}^{\text{tree}}} \,
\, = \,  \sum_{l=0}^\infty a^l \ \widehat \cA^{(l)}_{n} \, = \, \sum_{k=0}^{n-4} \sum_{l=0}^\infty a^l \ \widehat \cA^{(l)}_{n; k} \ . 
\end{equation}
Unlike the analogous correlator expansion~(\ref{corLeftloop}) all $l$ loop contributions come with $a^l$.

Our conjecture is roughly that ``the square of the amplitude is equal
to the correlation function in the light-like limit''. More concretely
then we write
\begin{equation}
\boxed{\lim_{x^2_{i \, i+1} \rightarrow \, 0} \, \sum_{l\ge 0} a^l \frac{G^{(l)}_n}{G^{\mathrm{tree}}_{n;0}} \,
 \, = \,
\Bigg( \sum_{l=0}^\infty a^l \ \widehat \cA^{(l)}_{n}\Bigg)^2}
\label{corConj2}
\end{equation}
in the planar limit, which is just a rewriting of
equation~(\ref{corConj1}) in the introduction without the coupling
dependent rescaling of
theta.

Note that although the right-hand side is simply the full superamplitude, the left-hand side is {\em{not}}  the correlator simply due to the fact that the powers of the coupling are not correct (see the discussion below (\ref{corLeftloop})) and this is why we write the explicit expansion on both sides. 
A similar issue arises~\cite{simon} when comparing the super Wilson loop  to amplitudes.

There are a few more ingredients  we need in order  to properly interpret this equation. Firstly, on the left-hand side the correlator is
defined in analytic superspace, with variables $x,y$ and $\rho$,
whereas on the right-hand side the variables are $x, \chi$. In order
to make sense of the equation we need to identify these
variables. We will find that the Grassmann variables are identified as
follows (a fact which follows straightforwardly from the known
expressions of both variables in terms of the standard Minkowski
superspace variable $\theta$ and is derived in appendix~\ref{sec:nairs-eta-as})
\begin{equation}
\chi_{i} \, = \, \la i| (\rho_i - \rho_{i \, i+1} \, y_{i \, i+1}^{-1} \, y_i) \, ,
\qquad
\chi_{i}' \, = \, \la i| \rho_{i \, i+1} \, y_{i \, i+1}^{-1} \, , \qquad
\la i| \, = \,  \lambda_{i}^{\alpha} \, .
\label{chiToRho}
\end{equation}
The labels in the last equation
exclusively indicate the point in superspace to which the variables belong.
We surpress the Lorentz and internal indices. They link up naturally if we keep their positions always as given in (\ref{defAna}) together with
$(y^{-1})_{a}{}^{a'}$. So for example $\rho_{i\,i+1} y_{i\,i+1}^{-1} y_i$ stands for $\rho_{i\,i+1}^{\alpha a} (y_{i\,i+1}^{-1})_{a}{}^{a'} y_{i\,a'}{}^{b}$ etc.
Further, we have split the $SU(4)$ index $A$ into its $SU(2)\times
SU(2)$ subgroup pieces, so $\chi^A = (\chi^a, \chi^{a'})$ which are in
turn denoted by $(\chi, \chi')$. 

The second thing we need to know is how to regularise,
since as it stands, both sides of the duality relation \p{corConj2} diverge. 
For generic $x_i$, all $n$-point functions of $\cT$ are
finite and (super)conformal order by order in perturbation theory.
The limit $x_{i \, i+1}^2 \rightarrow \, 0 \, , i \in \{1,\ldots,n\}$
(with the cyclic identification
$x_{n+1} = x_1$) puts the $n$ operators at the vertices of an $n$-gon with
light-like edges. In this limit, the correlators develop two kinds of
singularities. There are power singularities as for the tree-level correlator
\begin{equation}
G_{n ; 0}^{(0)}(1,\ldots,n)|_{\bar \rho_i =0} \, = \, {N^2_c-1 \over (4 \pi^2)^n} \, {y_{12}^2 \over x_{12}^2} \, {y_{23}^2 \over x_{23}^2} \,  \ldots {y_{n1}^2 \over x_{n1}^2} \, .
\label{treeNGon}
\end{equation}
In this formula we have displayed only the most singular term of the
connected tree, which turns out to be the highest power singularity also
in the loop corrections to $G_n$. Hence the ratio on the left-hand side of
(\ref{corConj2}) is free of power singularities. 

But the conformal loop
integrals found in the perturbative corrections to $G_n$ develop logarithmic 
divergences {when their external points become null separated}. These ``pseudo-conformal'' integrals require
regularisation.

This issue has already been encountered for the MHV duality \cite{EKS2}.
At the MHV one- and two-loop level (so $\bar \rho_i = \rho_i = \chi_i = 0$ and 
up to $O(a^2)$) our conjecture (\ref{corConj2}) yields
\begin{eqnarray}
\lim_{x^2_{i \, i+1} \rightarrow \, 0} \, \frac{G_{n;0}}{G_{n;0}^{(0)}} \,
\bigl(x_1,\ldots,x_n) & = &
1 + 2 \, a\, \widehat \cA_{n;0}^{(1)}(x_1,\ldots,x_n)  \\ && + \, 2 \, a^2  
\left(\widehat \cA_{n;0}^{(2)}(x_1, \ldots,x_n) + \frac{1}{2} (\widehat \cA_{n;0}^{(1)}
(x_1, \nonumber \ldots,x_n))^2 \right) + O(a^3)
\end{eqnarray}
where we have simply input the amplitude expansion (\ref{eq:2}) into the right-hand side of the conjecture~(\ref{corConj2}) and expanded the square. {We recall that $\widehat \cA_{n;0}^{(\ell)} $ stands here for $l-$loop correction to the ratio of $n-$particle MHV amplitude to its
tree-level expression. } 

This was demonstrated in \cite{EKS2} for all $n$-point one-loop MHV amplitudes
and the four- and five-point MHV two-loop amplitudes in a non-standard
regularisation scheme: the integrand of the loop level correlator was
evaluated in four dimensions and to regularise only the measure of the
integration over the insertion points was modified to $D = 4 - 2 \, \epsilon$
dimensions (with $\epsilon<0$). This non-standard $x$-space regularisation
precisely mimics the usual $p$-space infrared prescription for the amplitudes. 

But more is true: at the level of the integrands we can stay in exactly four
dimensions because we need not worry about singularities. Exact equivalence
holds for the integrands themselves, which was verified for the MHV 
five- and six-point one- and two-loop amplitudes and conjectured for all other
cases in \cite{EKS3}. 

So how can we unambiguously define an integrand for a loop level
correlator?
A crucial point that enables us to do so (and hence to compare with amplitude
integrands) is that 
loop corrections to such $n$-point correlators can be computed
by means of multiple Lagrangian insertions\footnote{As mentioned, all the $\cN = 4$  
results { in this paper} are actually derived from calculations with $\cN = 2$ superfields
\cite{GIKOS}, either
in this work or in the literature that we quote. The insertion
procedure that is actually used is differentiation with respect to  the coupling constant
in the $\cN = 2$ { harmonic superspace} formalism, whose essential details are briefly summarised
in Appendix \ref{sec:cal-n=2-superfields}. These $\cN=2$ results are then uplifted to $\cN =4$ analytic superspace.}$^{,}$\footnote{In this paper we do the Wick rotation before deriving Feynman
rules so that amongst other changes the factor $i^l$ disappears from the
corresponding formula in \cite{twin}.} \cite{whatExactly,ESS,EKS2,
EKS3} so that we have:
\begin{align}
\la \cT(1) \ldots \cT(n) \ra^{(l)} \, 
&= \, \frac{1}{l!} \int d\mu_{0_1}
\ldots d\mu_{0_l}  \la {\cal T}({0_1}) \ldots
{\cal T}({0_l}) \cT(1) \ldots \cT(n)  \ra^{(0)} \, , \label{insertIt}
\end{align}
where the bracketed superscript $(l)$ indicates that this is the $l$ loop
contribution  and where $d\mu :=
d^4x \, d^4\rho$. The second equality follows {from~\p{TT} and~(\ref{eq:12}), the Grassmann integral just picks the $\rho^4$ component of the superfield $\cT$.}  On the right-hand side the integrand is itself a correlator
and furthermore a Born level correlator. Therefore this Born level
correlator provides an unambiguous
definition of the integrand  which we can  compare with the integrands coming, for example from the
amplitude  integrand results of~\cite{nima1,nima2}.  

Further rewriting this in terms of the $\rho^{4k}$ expansion terms in~(\ref{corLeftloop}) we thus have that
\begin{equation}\label{eq:3}
G_{n; k}^{(l)}(1,\ldots,n) \, = \, \frac{1}{l!} \int d\mu_{0_1} \dots d\mu_{0_l} \,
G_{(n+l); (k+l)}^{(0)}({0_1},\ldots, {0_l};1,\ldots, n)\ , 
\end{equation}
{where the semicolon after $0_l$ distinguishes the loop
integration variables from the outer points. On the right-hand side we have the same type of object $G_{n;k}$ as on
the left-hand side, but at tree level {and at a higher Grassmann level}. {However, in the light-cone limit the points $x_i$ (with $i=1,\ldots,n$) form a light-like polygon while the points $x_{0_k}$ (with $k=1,\ldots, \ell$) remain in arbitrary positions. }} 

So in summary all loop-level integrands of correlation functions can
be written in terms of tree-level higher point correlation
functions and hence via the duality the integrand of any amplitude at
any loop order can be
obtained from tree-level stress-tensor multiplet correlators.

For example,  at one and two loops, we have the MHV amplitude/correlator duality
\begin{equation}
\lim_{x^2_{i \, i+1} \rightarrow \, 0} \,
\frac{G_{n;0}^{(1)}}{G_{n;0}^{(0)}} \, \bigl(1,\ldots,n)=
\int d^4 x_0 d^4\rho_0 \lim_{x^2_{i \, i+1} \rightarrow \, 0} \, \frac{G_{n+1;1}^{(0)}}{G^{(0)}_{n;0}} \, \bigl(0;1,\ldots,n) \, = \, 2 \,
\widehat \cA_{n;0}^{(1)}(1,\ldots,n) \,,
\label{oneLoopMHV}
\end{equation}
which we interpret as the  integrand identity
\begin{equation}
 \int d^4\rho_0 \lim_{x^2_{i \, i+1} \rightarrow \, 0} \,
\frac{G_{n+1;1}^{(0)}}{G^{(0)}_{n;0}} \, \bigl(0;1,\ldots,n) \, = \, 2 \,
\widehat A_{n+1;0}^{(1)}(0;1,\ldots,n)\ . \label{oneLoopMHV}
\end{equation}
Here the integrand of the amplitudes (divided by the tree-level MHV
amplitude) is denoted by $\widehat A$ with the integration
points included in the list of arguments before the semicolon, whereas the
corresponding integral is denoted by $\widehat \cA$; so for example we have
\begin{equation}   
\widehat \cA^{(1)}_{n;0}(x_1,\ldots,x_n) \, = \, \int d^4x_0 \, \widehat A^{(1)}_{n+1;0}(x_0;
x_1,\ldots,x_n)\, .
\end{equation}

Similarly at two loops we have the integrand identity
\begin{eqnarray}
&& \frac{1}{2} \int d^4\rho_0 \, d^4\rho_{0'} \lim_{x^2_{i \, i+1} \rightarrow \, 0} \,
\frac{G_{n+2;2}^{(0)}}{G^{(0)}_{n;0}} \, \bigl(0,{0'};1,\ldots,n)
\label{twoLoopMHV} \\
&&  \ = \, 2 \left( \widehat A^{(2)}_{n+2;0}(x_0,x_{0'};x_1,\ldots,x_n) +
\frac{1}{2} \, \widehat A^{(1)}_{n+1;0}(x_0;x_1,\ldots,x_n)  \, \widehat A^{(1)}_{n+1;0}(x_{0'};x_1,\ldots,x_n)
\right)\ . \nonumber
\end{eqnarray}

In (\ref{oneLoopMHV}) we have used the Lagrangian component of an additional
$\cT(0)$ operator at point 0 to obtain the one-loop correction to the $n$-point
$O(\rho^0)$ correlator. The outer points were put onto a light-like $n$-gon
while the insertion point is integrated out. On the other hand, before
integration and without any light-like limit this is, of course, just a specific
Grassmann component of an $(n+1)$-point function of $\cT$'s. Then
according to the duality~(\ref{corConj2}), we can take this same
component of the correlator in an $(n+1)$-gon
limit to obtain the
$(n+1)$-point NMHV tree-level amplitude~\cite{twin}. Once again, this correspondence
holds at  the level of the integrands.
In the same way, the $O(\rho^8)$ part of an $(n+2)$-point function of
$\cT$'s can yield
\begin{itemize}
\item the two-loop $n$-point MHV amplitude, if two points are treated as
insertions and integrated out while the others are put onto an $n$-gon with
light-like edges. This is the situation in equation (\ref{twoLoopMHV}).
\item the one-loop $(n+1)$-point NMHV amplitude, if one point is treated as an
insertion and integrated out, while the others are put onto an $(n+1)$-gon.
\item the tree-level $(n+2)$-point NNMHV amplitude in an $(n+2)$-gon limit
without any integrations.
\end{itemize}
The possibility of obtaining various amplitudes from the same generating
object is reminiscent of the supersymmetric Wilson loop of \cite{simon}.

{ In the rest of the paper we provide a number of explicit examples of the duality \p{corConj2}, at tree and at loop level.}

\section{Five-point one-loop NMHV and six-point tree\\ NNMHV}
\label{useESS}
\begin{figure}[h!]
\psfrag{O}[cc][cc]{$\scriptstyle \cO$ }
\psfrag{L}[cc][cc]{$\scriptstyle \cL$ }
\psfrag{t1}[cc][cc]{$\scriptstyle  {\rm MHV}_4^{(1)}$ }
\psfrag{t2}[cc][cc]{$\scriptstyle  {\rm NMHV}_5^{(0)}$ }
\psfrag{t3}[cc][cc]{$\scriptstyle  {\rm MHV}_4^{(2)}$ }
\psfrag{t4}[cc][cc]{$\scriptstyle  {\rm NMHV}_5^{(1)}$ }
\psfrag{t5}[cc][cc]{$\scriptstyle  {\rm NNMHV}_6^{(0)}$ }
\psfrag{Eq1}[cc][cc]{$ G_{5;1}^{(0)}\quad \to $ }
\psfrag{Eq2}[cc][cc]{$ G_{6;2}^{(0)}\quad \to $ }
  \centering
  \includegraphics[height=7.5cm]{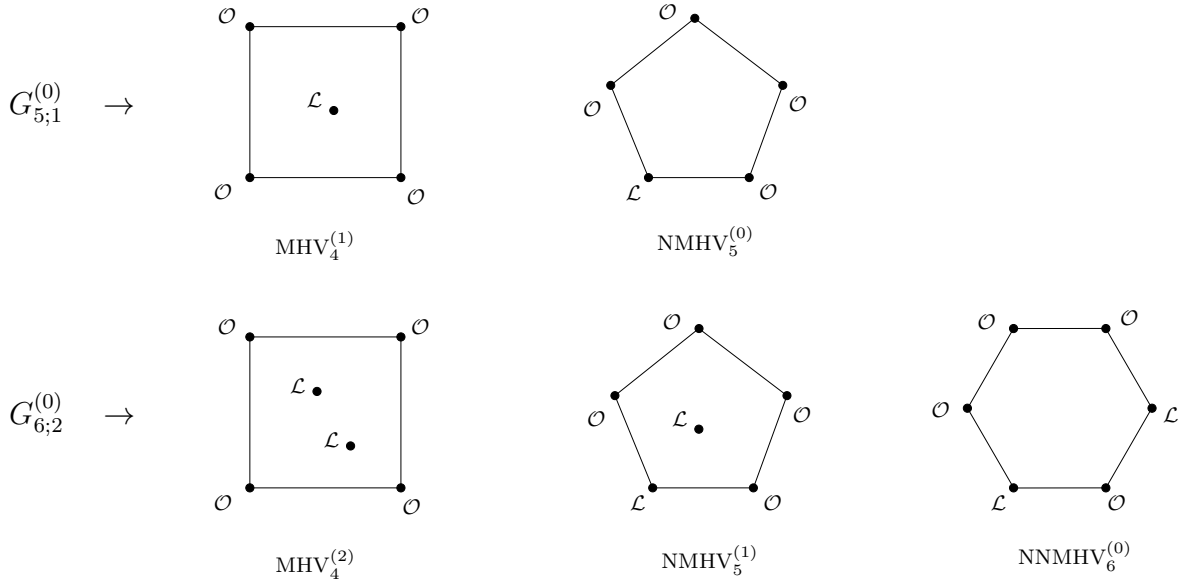}
  \caption{The different light-cone limits taken for the
points of the correlators $\vev{\cO\cO\cO\cO\cL}$ and $\vev{\cO\cO\cO\cO\cL\cL}$. Operators at neighbouring vertices of a polygon are
light-like separated, whereas those inside the polygon are located at arbitrary
points. }
  \label{fig:5point} 
\end{figure}

In this section we explore the one- and two-loop corrections to the simplest
correlator of the lowest-dimension components  $\cO = \tr(\phi^2)$ of the stress-tensor multiplets (see \p{TT}),
the purely bosonic correlator $G_{4;0} = \vev{\cO(1) \ldots \cO(4)}$.  We show that
the loop corrections to this four-point correlator, interpreted as Lagrangian insertions
\cite{whatExactly,ESS} (see (\ref{eq:3})), can give rise to several superamplitudes. So the integrand of the one-loop four point correlator $ G_{4;0}^{(1)} = \int d\mu_5 \, G_{5;1}^{(0)}$ and the integrand of the two-loop four-point correlator $ G_{4;0}^{(2)} = \frac{1}{2} \int d\mu_5 \, d\mu_6 \, G_{6;2}^{(0)}$ yield
the following amplitudes (see Fig.~\ref{fig:5point}):
 \begin{align}\label{}
& G_{5;1}^{(0)}\quad
\rightarrow \quad \left\{ \begin{array}{l}
  \mbox{\rm MHV}_4^{(1)}  \\
  \mbox{NMHV}_5^{(0)} \qquad \mbox{\phantom{N}Section~\ref{dscsa}}
\end{array}
  \right.  \nt
& G_{6;2}^{(0)}\quad
\rightarrow \quad \left\{ \begin{array}{l}
  \mbox{\rm MHV}_4^{(2)}  \\
 \mbox{NMHV}_5^{(1)} \qquad \mbox{\phantom{N}Section~\ref{dscsb}}  \\
  \mbox{NNMHV}_6^{(0)} \qquad \mbox{Section~\ref{dscsc}}  
\end{array}
  \right.   
\end{align}
Which amplitude is realised depends on how many $\cT$ operators are placed
on an $n$-gon with light-like edges, with the others treated as Lagrangian
insertions and integrated out. 

\subsection{Loop corrections to the four-point correlator  $G_{4;0}$} 

We now describe the loop corrections to the four-point correlators and
the related integrands (themselves higher-point tree-level
correlators) and in 
later subsections we will relate them in various light-like limits to
the respective amplitudes.

Two- and three-point functions of stress-tensor multiplets $\cT$'s do not
receive quantum corrections \cite{EHW}. The simplest non-trivial object to
study is thus indeed $G_{4;0}$. The form of its loop corrections is heavily
restricted by $\cN=4$ superconformal symmetry. This ``partial
non-renormalisation'' \cite{partialNonRen}, which we review in
appendix \ref{sec:part-non-renorm}
allows a remarkably simple writing of these loop corrections, and of
the related higher point Born level correlators. Here we simply
display the result of the computations originally done in~\cite{me1,simp,schalm,ESS,Bianchi}.

The one-loop four-point
correlator is 
given as the integral of a certain five-point correlator 
whereas the two-loop four-point correlator is the integral of a one-loop
five-point correlator, or alternatively of a Born level six-point correlator
 as follows:
\begin{align}
  \label{eq:6}
G_{4;0}^{(1)}(1,2,3,4)&=\int d^4 x_5 d^4 \rho_5 \,
G_{5;1}^{(0)}(1,2,3,4,5)\\\notag
G_{4;0}^{(2)}(1,2,3,4)&= \int d^4 x_5 d^4 \rho_5  G_{5;1}^{(1)}(1,2,3,4,5) 
\\ &
= \frac{1}{2} \int d^4 x_5 d^4 \rho_5 \int d^4 x_6 d^4
\rho_6 \,
G_{6;2}^{(0)}(1,2,3,4,5,6)\ .\label{eq:10}
\end{align}
The integrands themselves are
given by the simple formulae (see appendix~\ref{sec:part-non-renorm})
\begin{align}
  \label{eq:7}
G_{5;1}^{(0)}(1,2,3,4,5)|_{(\rho_5)^4}& = \frac{2 ( N^2_c-1)}{(4 \pi^2)^{5}} \times I \times (\rho_5)^4 {x_{13}^2
    x_{24}^2 \over x_{15}^2x_{25}^2x_{35}^2x_{45}^2}\\
  G_{6;2}^{(0)}(1,2,3,4,5,6)|_{(\rho_5)^4(\rho_6)^4}& = \frac{4 (N^2_c-1)}{(4 \,
\pi^2)^{6}} \times I \times
  (\rho_5)^4(\rho_6)^4 
\notag  \\
  & \times
  {x_{13}^2x_{24}^2}{\frac1{96}\sum_\sigma x_{\sigma(1)
      \sigma(2)}^2 x_{\sigma(3) \sigma(4)}^2x_{\sigma(5) \sigma(6)}^2
  \over x_{15}^2x_{25}^2x_{35}^2x_{45}^2
    x_{56}^2 x_{16}^2x_{26}^2x_{36}^2x_{46}^2}\label{eq:9}
\end{align}
where $I$ is the rational prefactor (this universal prefactor is a
consequence of  superconformal symmetry described further in
appendix~\ref{sec:part-non-renorm}) 
\begin{align}\label{eq:20}
 I &= \frac{y^2_{12}y^2_{23}y^2_{34}y^2_{41}}{x^2_{12}x^2_{23}x^2_{34} x^2_{41}}(1-s-t)
+\frac{ y^2_{12}y^2_{13}y^2_{24}y^2_{34}}{x^2_{12}x^2_{13}x^2_{24} x^2_{34}}(t-s-1)  \nt
& +\frac{y^2_{13}y^2_{14}y^2_{23}y^2_{24}}{x^2_{13}x^2_{14}x^2_{23} x^2_{24}}(s-t-1)+
\frac{y^4_{12} y^4_{34}}{x^4_{12}x^4_{34}}s  + \frac{y^4_{13} y^4_{24}}{x^4_{13}
x^4_{24}} + \frac{y^4_{14} y^4_{23}}{x^4_{14}x^4_{23}}t 
\end{align}
where we have introduced the conformal cross-ratios
\begin{align}\label{}
s=\frac{x^2_{12}x^2_{34}}{x^2_{13}x^2_{24}}\,, \qquad
t=\frac{x^2_{14}x^2_{23}}{x^2_{13}x^2_{24}}\, .
\end{align}

Note that the  $x$-space factor in the expression for $G_{5;1}^{(0)}$
is simply the one-loop box integrand, whereas the $x$ terms in
$G_{6;2}^{(0)}$  arise
from the two-loop ladder and one-loop box squared terms in
$F^{(2)}$~(\ref{4.10}). The sum in (\ref{eq:9}) is over all
permutations $\sigma$ of points 1 to 6. There is a 48-fold redundancy in
writing it like this since there are only 15 different terms in
the sum, so we divide by 48 in order to account for this; the remaining factor
1/2 adjusts the normalisation to meet the result (\ref{4.10}). 

Amplitude integrands can be obtained by taking different light-like
limits of these correlation functions as we now investigate.

\subsection{The $G_{4;0}$ $\leftrightarrow$  MHV$_4$ duality}

In this subsection we merely reproduce one of the results of \cite{EKS2} as
an illustration of the general procedure. The one- and two-loop corrections to
$G_4$ are given in (\ref{eq:6}), (\ref{eq:10}). To compare with four-point MHV amplitudes we need to put the four points of this 
correlator on the light-like square $x^2_{12}=x^2_{23}=x^2_{34}=x^2_{41}=0$,
which amongst other things, creates pole singularities in the prefactor $I$~(\ref{eq:20}). According to the 
duality \p{corConj2} we need to divide the correlator by the
connected tree-level correlator in order to remove these poles,
i.e. by
\begin{align} \label{}
G^{(0)}_{4;0 } = \frac{N^2_c-1}{(4 \pi^2)^4} \frac{y^2_{12} y^2_{23} y^2_{34}
y^2_{41}}{x^2_{12} x^2_{23} x^2_{34} x^2_{41}} + \mbox{subleading} \, .
\end{align}
Remarkably, this is equal to the leading singularity in the
prefactor $I(x_1,\ldots,y_4)$ (up to the factor $(N^2_c-1) / (4 \, \pi^2)^4$)  in
the light-like limit,  so that we obtain from (\ref{eq:6}-\ref{eq:9})
\begin{align}\label{}
\lim_{x^2_{i,i+1} \to \, 0} \frac{G_{4;0}}{G^{(0)}_{4;0}} &=1+\lim_{x^2_{i,i+1} \to \, 0} \left( a \frac{G_{4;0}^{(1)}}{G^{(0)}_{4;0}}+ a^2\frac{G_{4;0}^{(2)}}{G^{(0)}_{4;0}}\right) + O(a^3)\,\\
&=\lim_{x^2_{i,i+1} \to \, 0}  \Bigg[1+    \frac{2\, a}{16 \pi^2}
\int d^4 x_5 {x_{13}^2
    x_{24}^2 \over x_{15}^2x_{25}^2x_{35}^2x_{45}^2} \\
&\phantom{\lim_{x^2_{i,i+1} \to \, 0} \Big(}
+   \frac{2 \, a^2}{(16 \pi^2)^2} \int d^4 x_5 \, d^4 x_6 \left( 
{x_{13}^2x_{24}^2}{\frac1{96}\sum_\sigma x_{\sigma(1)
      \sigma(2)}^2 x_{\sigma(3) \sigma(4)}^2x_{\sigma(5) \sigma(6)}^2
  \over  x_{15}^2x_{25}^2x_{35}^2x_{45}^2
    x_{56}^2 x_{16}^2x_{26}^2x_{36}^2x_{46}^2}
\right) \Bigg].\nonumber
\end{align} 
Here in the light-like limit at one loop we immediately recognise the massless one-loop box function, whereas at two loops, in the sum over permutations, many terms are subleading in the light-like limit, and  we are left with the massless one-loop box squared together with the massless two-loop ladder diagram.  
The integrands occurring here exactly match the integrands 
of the one- and two-loop MHV$_4$ amplitudes~\cite{bern41,bern42} on taking the square.  

\subsection{The  $G_{5;1}^{(0)}$   $\leftrightarrow$ NMHV$_5^{(0)}$ duality} \label{dscsa} 

The simplest non-trivial example of the duality for tree-level amplitudes
concerns the five-point NMHV case. Note that this is the anti-MHV amplitude which is related to the MHV amplitude by parity. In the correlator picture
parity symmetry (in the sense of the scattered particles, not in the sense of
the fields) is far from obvious and thus even this case is quite a non-trivial
check of the duality.

According to the duality conjecture, to reproduce the NMHV 5-point tree-level correlator, we need to take the correlator $G_{5;1}^{(0)}$ and put all five points on the light-cone.

This should be compared with the previous subsection. There we were taking the same correlator $G_{5;1}^{(0)}$ but thinking of it as the integrand of the 
four-point one-loop correlator.  The integration point $x_5$ was thus in an
arbitrary position and we reproduced the four-point one-loop MHV amplitude (essentially the massless  box function).

Now, $x_5$ has become the fifth point of a light-like
pentagon, together with the external points $x_1,\ldots,x_4$. In other words,
in the new light-cone limit $x^2_{45}=x^2_{51}=0$, while $x^2_{41} \neq 0$. As
before, the light-cone poles are compensated by dividing out the tree-level
correlator
\begin{align}\label{tree5div}
G^{(0)}_{5;0 } = \frac{N^2_c-1}{(4 \pi^2)^{5}} \frac{y^2_{12}y^2_{23}y^2_{34}
y^2_{45}y^2_{51}}{x^2_{12}x^2_{23}x^2_{34} x^2_{45}x^2_{51}} + \mbox{subleading}\,.
\end{align}
One can easily check from its expression (\ref{eq:7}) that
\begin{align}\label{4.18}
\lim_{x^2_{i,i+1} \to \, 0} \left. \frac{G^{(0)}_{5;1}}{G^{(0)}_{5;0}} \,
\right\vert_{ \rho_5^4}= \, 2 \, \frac{x^2_{13} x^2_{24}}
{x^2_{14}x^2_{25}x^2_{35}} \, \frac{y^2_{41}}{y^2_{45}y^2_{51}} \, \rho_5^4 \, .
\end{align}  

Let us now compare this to the five-point tree-level amplitude
\begin{equation}
\frac{\cA^{(0)}_{5 }}{\cA^{\text{tree}}_{5 ; \text{MHV}}} \, = \, 
1+ \widehat \cA_{5;1}^{(0)}(1,\ldots,5) \, =
\, 1 + R_{12345} \, .
\label{eq:1}
\end{equation}
The R invariant \cite{annecySuperspace} on the right-hand side of
\p{eq:1} corresponds to the NMHV tree-level.  A general
expression for any R invariant in terms of momentum supertwistors 
was given in \cite{marksAndSparks}. The case $R_{12345}|_{\rho_5^4}$ is
evaluated in Appendix~\ref{sec:rho_i4-components-r} using the relation \p{chiToRho} between the $\chi$ and
$\rho$ variables. Here we merely state the result \p{resR135}:
\begin{equation}
R_{12345}|_{\rho_5^4} \, = \,
\frac{x^2_{13} x^2_{24}}{x^2_{14}x^2_{25}x^2_{35}}\frac{y^2_{41}}{y^2_{45}y^2_{51}} \
\rho_5^4 \, . \nonumber
\end{equation}
Finally, we compare with \p{4.18} finding
\begin{equation}
\lim_{x^2_{i,i+1} \to \, 0} \left. \frac{G^{(0)}_{5;1}}{G_{5;0}^{(0)}} \,
\right\vert_{ \rho_5^4} \, = \, 2  \, \left.\frac{\cA^{(0)}_{5;1}}{\cA^{\text{tree}}_{5 ; \text{MHV}}}\right\vert_{\rho_5^4}
\end{equation}
in perfect agreement with the conjectured duality relation \p{corConj2}.
The combined $Q, \bar S$ supersymmetries are powerful enough to
restore the full dependence on the left-handed Grassmann coordinates
$\rho_1,\ldots,\rho_4$.

\subsection{The  $G^{(1)}_{5;1}$ $\leftrightarrow$  NMHV$_5^{(1)}$ duality} \label{dscsb}

As shown in equation~(\ref{eq:10}) the integrand of the one-loop five-point
correlator $G_{5;1}^{(1)}$ in the gauge $\rho_1 =\rho_2=\rho_3  = \rho_4 = 0$  is given by the 6-point tree-level correlator
$G_{6;2}^{(0)}$.

Next, we take the pentagon light-cone limit $x^2_{12}= x^2_{23} = x^2_{34} =x^2_{45}= x^2_{51}=0$
of this correlator $G_{6;2}^{(0)}$, given in (\ref{eq:9}) and divide out the free
correlator \p{tree5div}. The result is
\begin{align}
\lim_{x^2_{i,i+1} \to \, 0} \left. \frac{G_{5;1}^{(1)}}{G^{(0)}_{5;0}}
\right\vert_{\rho_5^4} & = 
\lim_{x^2_{i,i+1} \to \, 0} \left. \int d^4 x_6 d^4 \rho_6 \frac{G_{6;2}^{(0)}}
{G^{(0)}_{5;0}}\right\vert_{\rho_5^4 \rho_6^4}   \nonumber \\
&=4\int d^4 x_6
    \, \frac{x^2_{13}x^2_{24}}
{x^2_{14}x^2_{25}x^2_{35}}  \ \frac{y^2_{41}}{y^2_{45}y^2_{51}} \  \rho_5^4 \  \left[ \frac{1}{2} \, g(6;1,2,3,4)  \, +  ({\rm cyclic} ) \right]
\nonumber \\
& = 4 \ \widehat \cA_{5;1}^{(0)}\ \widehat \cA_{5;0}^{(1)}|_{\text{even}}  
\phantom{\frac{1}{2}} \label{okN51}
\end{align} 
where ``$+$ (cyclic)''
means we add 4 terms obtained by cycling the points $1,\dots,5$ and
\begin{equation}
g(6;1,2,3,4)= {1 \over 4 \pi^2} {x^2_{13}x^2_{24} \over x^2_{16} x^2_{26} x^2_{36} x^2_{46}}
\end{equation}
is the one-loop box integrand with integration point $x_6$.
The third  line in \p{okN51} follows because the sum of box integrands $g(6;i,j,k,l)$ in the
second line is the same as the integrand of the {\em even part} of the
five-point one-loop MHV amplitude 
\cite{bern51} and hence also the same as the even part of the  integrand
$\widehat A_{5;0}^{(1)}$ of the one-loop NMHV${}_5$ $=
\overline{\text{MHV}}{}_5$ amplitude.  {Note that although this formula
is displayed as an integral identity, we really mean the identity of
the corresponding integrands.}

Does this agree with our duality conjecture \p{corConj2}? Expanding out the duality relation, we predict 
\begin{equation}\label{eq:11} 
\lim_{x_{i,i+1}^2 \rightarrow \, 0} \frac{G_{5;1}^{(1)}}{G_{5;0}^{(0)}}(1,\ldots,5)
 \, = \, 2 \, \Bigl( \widehat \cA_{5;1}^{(0)}(1,\ldots,5) \,
\widehat \cA_{5;0}^{(1)}(1,\ldots,5) \, + \, \widehat \cA_{5;1}^{(1)}(1,\ldots,5) \Bigr)\ .
\end{equation}
On the other hand, since at five points the NMHV amplitude is in fact an
$\overline{\text{MHV}}$ amplitude, it is equal to the tree level NMHV
amplitude multiplied by the complex conjugate of the MHV one-loop ratio.
Under complex conjugations the even part is invariant, but the parity
odd part gets a minus sign. We have therefore that $\widehat \cA_{5;1}^{(1)}=
\widehat \cA_{5;1}^{(0)} \left(\widehat \cA_{5;0}^{(1)}|_{\text{even}} -
  \widehat \cA_{5;0}^{(1)}|_{\text{odd}} \right)$. Then in the sum of terms
in~(\ref{eq:11}) the parity odd terms cancel and the prediction is in
precise agreement with what we find in \p{okN51}. For completeness, we display
the conjecture \cite{nima1,nima2} for the integrand $\widehat A_{5;1}^{(1)}$
in terms of momentum twistors at the end of Section \ref{numSec}. It does
indeed satisfy this conjugacy property.

\subsection{The  $G_{6;2}^{(0)}$ $\leftrightarrow$  NNMHV$_6^{(0)}$ duality} \label{dscsc}

This is a particular case of the duality between $\overline{\rm MHV}$
amplitudes and the maximally nilpotent part of correlators.
Due to the simple fact that we are taking the square of the amplitude
there are (NMHV tree)$\times$(NMHV tree) terms in addition to the NNMHV tree
part. In this respect the example is similar to the last one, but it is
interesting in its own right since the two contributions are distinct.

The maximally nilpotent part of the superamplitude/supercorrelator
duality conjecture \eqref{corConj2} yields the prediction:
\begin{align}
  \label{eq:19}
  \lim_{x^2_{i,i+1} \to \, 0} \frac{G_{n;  n-4}}{G^{\text{tree}}_{n;0}} \, = \,  
   \sum_{k=0}^{n-4} \ \widehat \cA_{n;{ k}} \ \widehat \cA_{n;{ n-4-k}} \ .
\end{align}
In particular, at Born level and six points we expect to find
\begin{align}\label{4.23}
\lim_{x^2_{i,i+1} \to \, 0}  \frac{G^{(0)}_{6;2}}{G^{(0)}_{6;0}}  &= \,
2 \, \widehat \cA^{(0)}_{6;2} \, + \, (\widehat \cA^{(0)}_{6;1 })^2 \nt
&= \, 2 \, \widehat \cA^{\text{tree}}_{6;\overline{\mathrm{\rm MHV}}} \, + \, 
   (\widehat \cA^{\text{tree}}_{6;\mathrm{NMHV} })^2 \,.
\end{align}
In order to check this we will first take the hexagon light-like limit of the correlator~(\ref{eq:9})
and then evaluate the $\overline{\mathrm{MHV}}$ and the additional
(NMHV)$^2$ part of the amplitude.

\subsubsection{The hexagon limit of the correlator  $G^{(0)}_{6;2}$}

The fully off-shell correlator $G^{(0)}_{6;2}$ is given in~(\ref{eq:9}). We have already seen how taking a four-point light-like limit of this leads to the two-loop four-point MHV integrand, and how taking the pentagon light-like limit leads to the five-point one-loop NMHV integrand. Now we wish to take the hexagon light-like limit in order to obtain the 6-point tree-level NNMHV amplitude. 

The hexagon limit creates new light-cone pole singularities, at
$x^2_{45}=x^2_{56}=x^2_{61}=0$. To cancel these we divide by the free correlator
 \begin{align}\label{}
 G^{(0)}_{6;0} = \frac{N^2_c-1}{(4 \pi^2)^{6}} \frac{y^2_{12}y^2_{23}y^2_{34}
y^2_{45}y^2_{56}y^2_{61}}{x^2_{12}x^2_{23}x^2_{34} x^2_{45}x^2_{56}x^2_{61}} +
\mbox{subleading}\,.
\end{align}
Only four terms (of the 15 in~(\ref{eq:9})) remain in the hexagon limit. 
The result is
\begin{align}\label{5.64}
&\lim_{x^2_{i,i+1} \to \, 0} \left. \frac{G^{(0)}_{6;2}}{G^{(0)}_{6;0}} \,
\right\vert_{\rho_5^4 \rho_6^4} \, = \, 2 \, {y_{14}^2 \over 
y^2_{45}y^2_{56} y^2_{61}  } \ \rho_5^4 \, \rho_6^4  \nt 
 &\qquad  \times   \frac{x_{13}^2 x_{24}^2}{x_{14}^2} \Big(
{x_{14}^2 \over x_{15}^2x_{35}^2 x_{26}^2 x_{46}^2 } +
{x_{13}^2 \over x_{15}^2x_{35}^2 x_{26}^2 x_{36}^2 } +
{x_{14}^2 \over x_{15}^2 x_{25}^2x_{36}^2 x_{46}^2 } +
{x_{24}^2 \over x_{25}^2x_{35}^2 x_{26}^2 x_{46}^2 }\Big)\,.
\end{align}
Next, we must compute both terms in the right-hand side of \p{4.23} in
analytic superspace and compare the result to \p{5.64}.

\subsubsection{Evaluating ${\widehat \cA}_{n;{\overline{\rm MHV}}}^{\rm tree}  $}
\label{sec:transl-r-invar-1}

Our first task is to find ${\widehat \cA}_{n;{\overline{\rm MHV}}}^{\rm
  tree}  $ and translate
it into analytic superspace.
The invariant ${\widehat \cA}_{n;{\overline{\rm MHV}}}^{\rm
  tree}  $
is simply the $\overline{\rm MHV}$ superamplitude divided
by the MHV superamplitude. In order to find this we will
employ the Nair $\eta$ variables \cite{Nair} rather than the $\chi$ variables
used in the rest of the text. These are related to the analytic
$\rho$ variables in Appendix~\ref{sec:nairs-eta-as}. We take a digression and
present a derivation valid for the $n$-point case although the explicit check
against the correlator will finally only be done for $n=6$.

The $\overline{\rm MHV}$ superamplitude can be written as
\begin{equation}
  \label{eq:21}
  { \cA}_{n;{\overline{\rm MHV}}}^{\rm
  tree}    = \delta^{(4)}( \sum_i \lambda_i \tilde \lambda_i)\ \delta^{(8)} \left(\sum_i \tilde \lambda_i
{\partial \over \partial \eta_i}\right) \prod_j {\eta_j^4 \over [j j+1]}\ .
\end{equation}
This form can be found by considering the standard form of the anti-MHV
superamplitude in terms of Fourier transformed Nair coordinates
$\tilde \eta$ and 
performing the explicit Fourier transform back to $\eta$'s.
We adopt the usual notation  
\begin{equation}
\la i j \ra \, = \, \lambda_i^\alpha \lambda_{j \, \alpha} \, , \qquad
[ i j ] \, = \, \tilde \lambda_i^{\dot \alpha} \tilde \lambda_{j \, \dot \alpha}
\, , \qquad x_{i \, i+1}^{\alpha \dot \alpha} = \lambda_i^\alpha \tilde
\lambda_i^{\dot \alpha}
\end{equation}
for products of the twistor variables parametrising
the light-like distances $x_{i \, i+1}$. In the two-component contractions we
do not introduce a weight factor of $1/2$, but we choose the normalisation
(see Eq.~\p{new} below)
\begin{equation}
\eta^4 \, = \, \frac{1}{4!} \, \epsilon_{ABCD} \, \eta^A \, \eta^B \, \eta^C
\, \eta^D \, = \, \frac{1}{4} \eta^2 \, \eta'^2 \label{normFourth}
\end{equation}
with $\eta^2 = \epsilon_{ab} \eta^a \eta^b$ and similar for $\eta'^2$.

We wish to consider \p{eq:21} in the gauge (on analytic superspace)
$\rho_1=\rho_2=\rho_3=\rho_4=0$. Using (\ref{genEtaSol}) this translates
into
\begin{equation}
  \label{eq:22}
  { \eta'_2}=\eta_2=0\qquad \eta_1+ \eta'_1 \, y_1=0 \qquad \eta_3+\eta'_3 \,
y_4 = 0\ .
\end{equation}
In this gauge we find that the $\overline{\rm MHV}$ superamplitude becomes
simply
\begin{equation}
{ \cA}_{n;{\overline{\rm MHV}}}^{\rm
  tree} =  \frac{1}{4} \, {[23]^2 [12]^2 y_{14}^2 (\eta'_1)^2
(\eta'_3)^2 \eta_4^4
\dots \eta_n^4 \over [12][23] \dots [n1]} \, = {\delta^{(8)}(\sum_i \lambda_i \,
\eta_i) \over \langle 12\rangle \langle23 \rangle \dots \langle n1 \rangle}
\times {\widehat  \cA}_{n;{\overline{\rm MHV}}}^{\rm
  tree} \ ,
\label{eq:23}
\end{equation}
and we have used  the definition of ${\widehat  \cA}_{n;{\overline{\rm MHV}}}^{\rm
  tree}$ in this gauge
as the aforementioned ratio. In the above we have split the four-component
$\eta^A$ into two two-component $\eta$ and $\eta'$, as previously.
We can use the delta function to eliminate two more $\eta^A$ from
${\widehat  \cA}_{n;{\overline{\rm MHV}}}^{\rm
  tree}$. We choose to eliminate $\eta_{n-1}$ and
$\eta_{n}$, after which writing $\delta^{(8)}(\sum_i \, \lambda_i \, \eta_i) = 
\eta_{n-1}^4 \, \eta_n^4 \, \langle n-1 \, n \rangle^4 + \dots$ yields 
a unique expression for ${\widehat  \cA}_{n;{\overline{\rm MHV}}}^{\rm
  tree}$:
\begin{equation}
  \label{eq:24}
  {\widehat  \cA}_{n;{\overline{\rm MHV}}}^{\rm
  tree}
 = {\langle 12\rangle \langle23 \rangle \dots \langle
n1 \rangle
    \over [12][23] \dots [n1]} \times {[12]^2 [23]^2 \over \langle n-1\, n
\rangle^4}
  \times y_{14}^2 \times (\eta'_1)^2 (\eta'_3)^2 \eta_4^4
\dots \eta_{n-2}^4
\end{equation}

Now that we have ${\widehat  \cA}_{n;{\overline{\rm MHV}}}^{\rm
  tree}$ in terms of $\eta,\eta'$, we just
need to re-express it in terms of the analytic $\rho$ variables putting in the
expression \p{genEtaSol} for $\eta(\rho),\eta'(\rho)$. We can start with
$\eta'_3$ and work upwards as follows:
\begin{eqnarray}
  \label{eq:25}
  \eta'_3 &=& {\langle 4| \rho_5 \, y_{45}^{-1} \over \langle 34 \rangle}\\
    \eta_4+\eta'_4 y_5&=&-{\langle 3| \rho_5 \over \langle34 \rangle}\\
\eta'_4&=& {\langle 5| \rho_6 \, y_{56}^{-1} \over \langle 54 \rangle}  +
O(\rho_5)\\
    \eta_5+\eta'_5 y_6&=&-{\langle 4| \rho_6 \over \langle 45 \rangle} +
O(\rho_5)\\ \eta'_5&=& {\langle 6| \rho_7 \, y_{67}^{-1} \over \langle 65
\rangle } +  O(\rho_5,\rho_6)\\
&\dots&
\end{eqnarray}
where the $O(\rho_5, \rho_6)$ terms indicate terms proportional to
$\rho_5$ or $\rho_6$. Such terms in ${\widehat  \cA}_{n;{\overline{\rm MHV}}}^{\rm
  tree}$ can be ignored:
For example, all possible occurrences of
$\rho_5$ are saturated by $(\eta'_3)^2 (\eta_4+\eta'_4 y_4)^2$. In the end
then we can safely substitute the following:
\begin{eqnarray}
  \label{eq:26}
  \eta'_{j-1}&=&{\langle j| \rho_{j+1} \, y_{j\,j+1}^{-1} \over \langle j-1\,j
\rangle}\qquad j=4 \dots n-2\\
  \eta_j+\eta'_j y_{j+1}&=&-{\langle j-1| \rho_{j+1} \over \langle j-1 \,j
       \rangle}\qquad j=4 \dots n-2\\
\eta'_{n-2}&=& {\langle n-1| \rho_{n} \, y_{n-1\,n}^{-1} \over \langle n-2\,n-1
  \rangle}\\
\eta'_1&=&{\langle n| \rho_n \, y_{n1}^{-1} \over \langle n1\rangle}
   \end{eqnarray}
which, using that
 \begin{equation}
\eta_j^4 = \frac{1}{4} \, {\eta'_j}^2 \eta_j^2 = \frac{1}{4} \, {\eta'_j}^2 (
 \eta_j+\eta'_j y_{j+1})^2\label{eq:27}
 \end{equation}
and\footnote{Once again this equation is valid only in the chain of
substitutions.} 
\begin{equation}
  \label{eq:28}
 \frac{1}{4} \, (\eta'_{j-1})^2
(\eta_j+\eta'_j y_{j+1})^2 = {\rho_{j+1}^4\over \langle j
    \,j-1\rangle^2 y_{j\,j+1}^2} \qquad j=4\dots n-2 
\end{equation}
gives us
\begin{equation}
  \label{eq:29}
(\eta'_1)^2 (\eta'_3)^2 \eta_4^4
\dots \eta_{n-2}^4 = {\rho_5^4 \dots \rho_n^4 \langle n-1 n \rangle^4 \over
\langle 34\rangle^2 \langle 45
  \rangle^2 \dots \langle n 1 \rangle^2 \, y_{45}^2 \, y_{56}^2
   \dots y_{1 n}^2}\ .
\end{equation}
Inserting this into (\ref{eq:24}) finally yields the $n$-point
$\overline{\rm MHV}$ 
invariant in analytic superspace 
\begin{equation}\label{5.41}
   {\widehat  \cA}_{n;{\overline{\rm MHV}}}^{\rm
  tree}= {x_{13}^4 x_{24}^4 \over x_{13}^2 x_{24}^2
\dots x_{n2}^2} \times
{y_{12}^2 y_{23}^2 y_{34}^2 y_{14}^2 \over y_{12}^2y_{23}^2
   \dots y_{1 n}^2} \times \rho_5^4 \dots \rho_n^4\ ,
\end{equation}
where we have simply relied on
$\la i \, i+1 \ra [i \, i+1] =  x^2_{i \, i+ 2}$. 

\subsubsection{Evaluating $(\widehat \cA^{\text{tree}}_{6;\mathrm{NMHV}
  })^2$ }

We now wish to compute the other contributions according to the
duality conjecture at this level (\ref{eq:19}). For simplicity we
concentrate on the six point case, and we wish to find the
contribution of $(\widehat \cA^{\text{tree}}_{6;\mathrm{NMHV} })^2$
to the correlation function according to the duality
conjecture (\ref{4.23}). At six points we have \cite{annecySuperspace}
\begin{align}\label{}
\widehat \cA^{\text{tree}}_{6;\mathrm{NMHV} }=
R_{61234}+R_{61245}+R_{62345}=R_5+R_3+R_1\ ,
\end{align}  
where in the third expression we are defining (for six points only) $R_i$
to be the invariant $R_{jklmn}$ which does not contain the index $i$.
Therefore  
\begin{equation}
  \label{eq:32}
 (\widehat \cA^{\text{tree}}_{6;\mathrm{NMHV} })^2 = (R_5+R_3+R_1)^2 =2
 R_{1} R_{3}+ 2 R_{1}R_5 + 2 R_{3}R_{5}\ .
\end{equation}
In this section --  like in the rest of the article --  we employ momentum
supertwistor variables. One of these is the $\chi_i$ parameter that we have
frequently mentioned. The second variable is a projective four-vector
$Z_i = (\lambda_i,\mu_i)$ with 
\begin{equation}
\mu_{i\,\dot \alpha} \, = \,  \, \lambda_i^\alpha (x_i)_{\a\dot \alpha}\ \,,
\end{equation}
or conversely
\begin{align}
(x_i)_{\a\da} =  \frac{
\lambda_{i \, \alpha} \, \mu_{i-1 \, \dot \alpha}-\lambda_{i-1 \, \alpha} \,
\mu_{i \, \dot \alpha}  }{\vev{i-1,i}} \label{makeX}
\end{align}
An $n$-point amplitude can be parametrised by a set $\{Z_1,\ldots,Z_n\}$
with the association \cite{hodges}
\begin{equation}
x_i \leftrightarrow (Z_i,Z_{i+1}) \, .
\end{equation}
The on-shell constraints $x_{i,i+1}^2 = 0$ are solved
by construction, which can be seen for example from the defining relation
for the four-bracket of twistors:
\begin{equation}
\la i \, j \, k \, l \ra \, = \, \mathrm{Det}\, (Z_i \, Z_j \, Z_k \, Z_l)
\, = \, \epsilon_{ABCD} Z_i^A  Z_j^B Z_k^C Z_l^D
\label{detForm}
\end{equation}
If the twistors pertain to two points $x_i, x_j$ this becomes 
\begin{equation}
\la i-1 \, i \; j-1 \, j \ra \, = \,
\la i-1 \, i \ra \, \la j-1 \, j \ra \, x^2_{ij} \, \label{detToProp}
\end{equation}
so that $x^2_{i,i+1} = 0$ due to the doubling of $Z_i$ in the determinant on
the left hand side.

Each $R_i$ has the form $R_{i}=c_{i} (\Sigma_{i})^4$ where $\Sigma$
is defined in (\ref{5.8}) in Appendix~\ref{sec:rho_i4-components-r} and $c$ is the bosonic factor of
$(\Sigma)^4$ from \p{NMH}. Hence, we find that the nilpotent pieces we
need to compute arise from terms like $(\Sigma_1)^4(\Sigma_3)^4$,
$(\Sigma_1)^4(\Sigma_5)^4$ and $(\Sigma_3)^4(\Sigma_5)^4$.  
As before, we want to use the gauge
\begin{align} \label{5.44}
\rho_1=\rho_2=\rho_3=\rho_4=0\ .
\end{align}
The identification between the $\chi$ and the $\rho$ parameters is
\begin{align}
  \label{chirho}
  \chi_i+ \chi'_i \, y_i = \bra{i} \rho_i \, , \qquad
\chi_i+\chi'_i \, y_{i+1}= \bra{i}\rho_{i+1}\ ,
 \end{align}
or inversely (cf. \p{chiToRho})
\begin{equation}
\chi_i \, = \bra{i}  (\rho_{i}- \rho_{i\, i+1} \, y_{i\,i+1}^{-1} \, y_i) \, , \qquad
\chi'_i \, = \bra{i}  \rho_{i\,i+1} \, y_{i\,i+1}^{-1}
 \ . \nonumber
\end{equation}
In terms of supertwistor variables the gauge \p{5.44} reads
\begin{equation}
  \chi_1^A=\chi_2^A =\chi_3^A = 0 \qquad \chi_4+ \chi'_4 \, y_4=0
\qquad \chi_6+\chi'_6 \, y_1 = 0\ .\label{eq:34}
\end{equation}
This gives us (all $\Sigma_i^A$ in the next formula carry a four-component
index)
\begin{align}
  \Sigma_{1 }=& \, \chi_{4 } \, \veve{5623} + \chi_{5 } \, \veve{6234} +
\chi_{6 } \, \veve{2345}\nt
 \Sigma_{3 }=& \, \chi_{4 } \, \veve{5612} + \chi_{5 } \, \veve{6124} +
\chi_{6 } \, \veve{1245}\nt
 \Sigma_{5 }=& \, \chi_{4 } \, \veve{6123}+\chi_{6 } \, \veve{1234}\nt
 \tilde \Sigma_{3 } := \Sigma_{3 }- {\veve{6124} \over \veve{6234}}
 \, \Sigma_{1 } =& \Big(\chi_{4 } \, {\veve{6123}\veve{6245} }+ \chi_{6 } \, 
 {\veve{2456}\veve{4123} } \Big)\veve{6234}^{-1}
\end{align}
where in the last equation we define a linear combination,
$\tilde \Sigma_{3}$ of $\Sigma_3$ and $\Sigma_1$ which is independent of
$\chi_5^A$ and we have simplified the expression using the twistor
identity
\begin{align}\label{twid}
  \veve{abij}\veve{abkl}+  \veve{abik}\veve{ablj}+
  \veve{abil}\veve{abjk}=0\ .
\end{align}

In our gauge, due to the relations (\ref{eq:34}) $\chi_{4}^A$ and
$\chi_{6}^A$ have only two independent components each, say, $\chi'_4$ and
$\chi'_6$, therefore $(\Sigma_{5})^4$ saturates all the $\chi_{4}^A $ and
$\chi_{6}^A$ variables giving (c.f. the derivation of \p{doubleDelta} in
Appendix~\ref{sec:rho_i4-components-r})
\begin{align}
  (\Sigma_{5})^4 =& \, \frac{1}{4} \,
\veve{6123}^2\veve{1234}^2 (\chi'_4)^2 (\chi'_6)^2 \, y_{14}^2\ .
\end{align}
So one need only consider
$\chi_5$ terms when multiplied by $\Sigma_5$. One then quickly finds
\begin{align}\notag
(\Sigma_1)^4  (\Sigma_5)^4 =& \, \frac{1}{4} \,
\veve{6123}^2\veve{1234}^2\veve{6234}^4
(\chi'_4)^2 (\chi'_6)^2 \chi_5^4 \, y_{14}^2 \, , \\ \notag
(\Sigma_3)^4  (\Sigma_5)^4 =& \, \frac{1}{4} \,
\veve{6123}^2\veve{1234}^2\veve{6124}^4
(\chi'_4)^2 (\chi'_6)^2 \chi_5^4 \, y_{14}^2 \, , \\
(\Sigma_1)^4  (\Sigma_3)^4 = \, (\Sigma_1)^4  (\tilde \Sigma_{3})^4=& \,
\frac{1}{4} \, 
\veve{6123}^2 \veve{1234}^2\veve{2456}^4 (\chi'_4)^2 (\chi'_6)^2 \chi_5^4
\, y_{14}^2 \, .
\label{eq:37}
\end{align}
Then we input the bosonic factors $c_{i}$  from \p{NMH}
 and also rewrite $\chi$ in terms of $\rho$ in our gauge,
 using (\ref{chirho}) to obtain
\begin{equation}
  \label{eq:38}
\frac{1}{4} (\chi'_4)^2 (\chi'_6)^2 \chi_5^4 \, y_{14}^2 = 
\, {\vev{45}^2 \vev{56}^2 } \times {y_{14}^2 \over y_{16}^2
y_{45}^2 y_{56}^2} \times \rho_5^4 \, \rho_6^4\ .
\end{equation}
Further, we use \p{detToProp} to rewrite some of the twistor factors in
terms of $x$ space variables and we find
\begin{align}
  \label{eq:43}
  R_{1}R_{3} &= \, {x_{13}^2x_{24}^2 \over x_{15}^2 x_{35}^2 x_{26}^2x_{46}^2}
\times { \veve{4562}^2 \veve{6123}\veve{1234} \over
  \veve{6124}\veve{2346}\veve{5623}\veve{4512}} \times {y_{14}^2 \over y_{16}^2
  y_{45}^2 y_{56}^2} \times \rho_5^4 \rho_6^4\nt
 R_{1}R_{5} &= \, {x_{13}^2x_{24}^2 \over x_{15}^2 x_{35}^2 x_{26}^2x_{46}^2}
\times { \veve{2346}^2 \veve{4561}\veve{5612} \over
  \veve{4562}\veve{6124}\veve{3461}\veve{2356}} \times {y_{14}^2 \over y_{16}^2
  y_{45}^2 y_{56}^2} \times \rho_5^4 \rho_6^4\nt
  R_{3}R_{5} &= \, {x_{13}^2x_{24}^2 \over x_{15}^2 x_{35}^2 x_{26}^2x_{46}^2}
\times { \veve{6124}^2 \veve{2345}\veve{3456} \over
  \veve{2346}\veve{4562}\veve{1245}\veve{6134}} \times {y_{14}^2 \over y_{16}^2
  y_{45}^2 y_{56}^2} \times \rho_5^4 \rho_6^4 \ .
\end{align}

The sum of these three terms can be simplified by rewriting the momentum
twistor conformal invariants in terms of six complex variables $z_i$ by using
the replacement \cite{spradGonch} 
\begin{align}
 \veve{ijkl}=z_{ij} z_{ik} z_{il}z_{jk} z_{jl}z_{kl}
\end{align}
where $z_{mn}=z_m -z_n$.~\footnote{In fact it is practically much more
  straightforward to make the replacement $\veve{ijkl} \rightarrow
  \epsilon_{ijklmn}z_{mn}$ which appears to give the same result for conformally invariant objects.} The advantage of these variables is that
identities such as \eqref{twid} become manifest, 
in this case $z_{wx} z_{tv}+z_{xv} z_{tw}+z_{vw}z_{tx}=0$.
In this way we get
\begin{align}
  R_{1}R_{3} &= {x_{13}^2x_{24}^2 \over x_{35}^2 x_{46}^2
    x_{15}^2x_{26}^2}
\times { z_{13}^2 z_{54} z_{56} \over
  z_{35}z_{51}z_{14}z_{36}}  \times {y_{14}^2 \over y_{16}^2
  y_{45}^2 y_{56}^2} \times \rho_5^4 \rho_6^4\nt
  R_{1}R_{5} &= {x_{13}^2x_{24}^2 \over x_{35}^2 x_{46}^2
    x_{15}^2x_{26}^2}
\times { z_{51}^2 z_{32} z_{34} \over
  z_{13}z_{35}z_{52}z_{14}} \times {y_{14}^2 \over y_{16}^2
  y_{45}^2 y_{56}^2} \times \rho_5^4 \rho_6^4\nonumber\\
  R_{3}R_{5} &= {x_{13}^2x_{24}^2 \over x_{35}^2 x_{46}^2
    x_{15}^2x_{26}^2}
\times { z_{35}^2 z_{16} z_{12} \over
  z_{51}z_{13}z_{36}z_{52}} \times {y_{14}^2 \over y_{16}^2
  y_{45}^2 y_{56}^2} \times \rho_5^4 \rho_6^4\ .\label{zsum}
\end{align}

Finally we are interested in the sum of these terms. It turns out that
although each term individually is complicated (at least when
expressed in $x$ space) the sum of terms has a very simple form.
We have that
\begin{align}
  \label{eq:45}
  { z_{51}^2 z_{32} z_{34} \over
  z_{13}z_{35}z_{52}z_{14}}+{ z_{35}^2 z_{16} z_{12} \over
  z_{51}z_{13}z_{36}z_{52}}+{ z_{13}^2 z_{54} z_{56} \over
  z_{35}z_{51}z_{14}z_{36}}&\equiv {z_{12}z_{45}\over z_{14}z_{25}}+
{z_{16}z_{34}\over z_{14}z_{36}}+{z_{23}z_{56}\over z_{25}z_{36}}\nt
&= {x_{13}^2 x_{46}^2 \over x_{36}^2 x_{14}^2} + {x_{35}^2 x_{26}^2 \over x_{36}^2 x_{25}^2}+{x_{15}^2 x_{24}^2 \over x_{14}^2 x_{25}^2}
\end{align}
where the first line is an algebraic identity, and in the second line
we have replaced the $z$ (via the momentum twistors) back with the
$x$'s. Remarkably all parity-odd pieces (which appear in the
R invariants themselves) completely cancel in this expression.

Putting this result (\ref{zsum}), \p{eq:45} first into \eqref{eq:32} and then
together with the expression \p{5.41} for ${\widehat  \cA}_{6;{\overline{\rm MHV}}}^{\rm
  tree}$,  we obtain
the right-hand side of the duality relation \p{4.23}:
\begin{align}
& 2 {\widehat  \cA}_{6;{\overline{\rm MHV}}}^{\rm
  tree} + 
   ({\widehat  \cA}_{6;{\rm NMHV}}^{\rm
  tree} )^2 =  \, 2 \, {y_{41}^2 \over  y^2_{45}y^2_{56} y^2_{61}  } \ 
 \rho_5^4 \, \rho_6^4  \nt 
 &\qquad  \times   \frac{x_{13}^2 x_{24}^2}{x_{14}^2} \Big({x_{14}^2 \over x_{15}^2
x_{35}^2 x_{26}^2 x_{46}^2 } + \Big[{x_{13}^2 \over x_{15}^2 x_{35}^2 x_{26}^2 x_{36}^2} +
{x_{14}^2 \over x_{15}^2x_{25}^2 x_{36}^2 x_{46}^2 }+{x_{24}^2 \over x_{25}^2x_{35}^2
x_{26}^2 x_{46}^2 }\Big]\Big)\ .\label{eq:16}
\end{align}
Remarkably this is in perfect agreement with the correlator prediction
\p{5.64}.

\section{The six-point one-loop NMHV amplitude} \label{newIntCheck}

In the previous  section we have illustrated how the off-shell
calculation \cite{ESS} of the tree-level correlator
$G_{6;2}^{(0)}=\la \cO(1) \, \cO(2) \, \cO(3) \, \cO(4) \, \cL(5) \, \cL(6)
\ra^{(0)} $ at $\rho_i = 0, \, i  \in \{1,\ldots,4\}$ can yield three
different amplitude integrands:
\begin{itemize}
\item \emph{MHV four-point two-loop amplitude} \\ in the square light-cone
limit $x_{i \, i+1}^2 \rightarrow 0, \, 1 \leq i \leq 4$, under the
double integral 
$\int d^4x_5 d^4\rho_5 \, d^4x_6 d^4\rho_6$.
\item \emph{NMHV five-point one-loop amplitude} \\ in the pentagon
light-cone limit $x_{i \, i+1}^2 \rightarrow 0, \, 1 \leq i \leq 5$ and
under $\int d^4x_6 d^4\rho_6$.
\item \emph{NNMHV six-point tree amplitude} \\ in the hexagon light-cone
limit $x_{i \, i+1}^2 \rightarrow 0, \, 1 \leq i \leq 6$ without any integration.
\end{itemize}
In~\cite{EKS2,EKS3},  the methods of \cite{ESS} were applied
to find 
$G_{7;2}^{(0)} = \la \cO(1) \ldots \cO(5) \, \cL(6) \, \cL(7)
\ra^{(0)}|_{\rho_6^4 \, \rho_7^4}$
and
$G_{8;2}^{(0)}|_{\rho_7^4 \, \rho_8^4} = \la \cO(1) \ldots \cO(6) \, \cL(7) \, \cL(8) \ra|_{ \rho_7^4 \, \rho_8^4}$
in order to demonstrate the duality
between these correlation functions put on the light-cone
and the MHV two-loop five- and six-point amplitudes.
The \emph{integrands} for the MHV amplitudes derived from the correlators
turned out to be equal to those predicted from BCFW recursion rules in
\cite{nima1}.

The cases studied in the last section already provide very
non-trivial evidence for the duality beyond MHV. However, to hopefully
remove any further
doubt, we here give an example which shows that the duality applies
simultaneously beyond both 
MHV/$\overline{\text{MHV}}$ sectors and beyond the tree-level sector and in
particular can
correctly relate the full integrands
even including a highly non-trivial parity odd piece.
To this end  we want to obtain the \emph{NMHV six-point one-loop
amplitude} from a new hexagon light-cone limit of the correlator
$G_{7;2}^{(0)}$. The pentagon light-like limit of this correlator was
already  found in~\cite{EKS2} to 
be dual to
the \emph{MHV five-point two-loop amplitude integrand}. 

Unlike the cases considered in previous sections, since there is
an additional 
outer point,  superconformal symmetry alone is not sufficiently
powerful to reconstruct 
the full ${\cal N} = 4$ correlator/amplitude from a single $\rho_i^4 \rho_j^4$
component. We therefore show how to do this reconstruction starting from
certain ${\cal N} = 2$ projections with five hypermultiplet bilinears and
one Lagrangian component. It is then enough to check for one Grassmann
component that the integrand as computed from the correlator is equal to the
momentum twistor expression in \cite{nima2}.
In the following subsection we start by building up some technology needed
to master the large parity-odd sector of the calculation in a manifestly
conformal way.

Obviously, according to our duality we could also construct the
\emph{NNMHV seven-point tree level} amplitude from this correlator $G_{7;2}^{(0)}$, although we refrain from doing so
because of the volume of that calculation. It is probably more striking
to see the duality at work at NMHV one-loop level in a non-trivial
case anyhow.

\subsection{An $x$-space toolkit for 6-point one-loop amplitudes}
\label{traceSec}

We wish eventually to construct the correlator $G_{7;2}^{(0)}$ and compare with
the six point NMHV 1-loop integrand. We therefore have a total of 7
points, the integration point (which we label 0) and the six outer
points which will be light-like separated.) We first develop some
techniques for writing down conformal invariants of this form in the
hexagon light-like limit.

The pseudo-conformal one-loop integrals that we will encounter at
six-points in the rest of
this section are the pentagons $p_i$ and boxes $g_{ij}$ defined as
\begin{eqnarray}
p_1 & = & \frac{1}{4 \pi^2} \int \frac{d^4x_0 \, x^2_{10}}{x^2_{20} x^2_{30}
x^2_{40} x^2_{50} x^2_{60}} \qquad \qquad g(1,2,3,4) \, = \, \frac{1}{4 \pi^2}
\int \frac{d^4x_0 \,
}{ x^2_{10} 
x^2_{20} x^2_{30} x^2_{40}}  , \\
g^{\mathrm{1m}}_{12} & = & g({3,4,5,6}) \, , \qquad
g^{\mathrm{2mh}}_{13} \, = \ g({2,4,5,6}) \, , \qquad
g^{\mathrm{2me}}_{14} \, = \ g({2,3,5,6}) \, . \phantom{\int} \nonumber
\end{eqnarray}
The labels indicate the factors which are missing from the maximal
denominator $x^2_{10} \ldots x^2_{60}$. Cyclic shifts yield six such integrals
in the first three cases, while there are only three two-mass easy boxes.
Thus we have a total of 21 integrals.

The hexagon light-cone limit $x^2_{i \, i+1} \rightarrow 0 , \, i \in
\{1,\ldots,6\}$ permits three finite cross ratios
\begin{equation}
u_1 \, = \, \frac{x^2_{13} x^2_{46}}{x^2_{14} x^2_{36}} \, , \qquad
u_2 \, = \, \frac{x^2_{15} x^2_{24}}{x^2_{14} x^2_{25}} \, , \qquad
u_3 \, = \, \frac{x^2_{26} x^2_{35}}{x^2_{25} x^2_{36}} \, .
\end{equation}
This definition puts those $x^2_{ij}$ into the denominator in which the
points are at opposite corners of the hexagon. Cyclic shifts along the hexagon
therefore permute these $u$'s but do not invert them.

On several occasions we find the following same fixed combinations of the
21 scalar integrals with polynomials of $u_1,u_2,u_3$
making an appearance. It is therefore
convenient to introduce these combinations as a second basis for the 1
loop integrals:
\begin{eqnarray}
\tilde p_1 & = & (1-u_3) \, \frac{x^2_{24} x^2_{35} x^2_{46}}{x^2_{14}} \; p_1 \, ,
\nonumber \\
\tilde g_{12} & = & (1-u_1+u_2-u_3) \, x^2_{35} x^2_{46} \; g^{\mathrm{1m}}_{12} \, ,
\nonumber \\
\tilde g_{13} & = & (1-u_1-u_2-u_3+2 u_2 u_3) \, x^2_{25} x^2_{46} \;
g^{\mathrm{2mh}}_{13} \, , \\
\tilde g_{14} & = & (1-u_3)(1-u_1-u_2-u_3) \, x^2_{25} x^2_{36} \;
g^{\mathrm{2me}}_{14} \, , \nonumber
\end{eqnarray}
with the 17 others defined in the obvious way by cyclic shifts. 

There is an interesting \emph{integrand identity} involving these combinations
with coefficients $\pm 1$ only
\begin{equation}
0 \, = \, \sum_{i=1}^6 \left(- \, \tilde p_i \, + \, \tilde g_{i \, i+1} \, - \,
\tilde g_{i \, i+2} \right) \, + \, \sum_{i=1}^3 
\, \tilde g_{i \, i+3} \, . \label{wow}
\end{equation}
Putting all terms under a common integral over $x_0$ and factorising produces
a numerator polynomial with 87 terms, all composed of seven $x^2_{ij}$ factors.
This is of conformal form; every term has weight two at all points. Upon
substituting rational numbers we may verify that the polynomial
identically vanishes in the hexagon light-cone limit, regardless of the choice
of $x_0$. This could in principle be shown by expanding the Lorentz
invariant $x^2 = x_0^2 - x_1^2 - x_2^2 - x_3^2$ in its components, although in
practice this is hardly possible by the sheer size of the problem. It
can also be shown in momentum twistors and is presumably related to
the Gram determinant.

If a regulator is introduced, the sum of integrals in (\ref{wow}) should not
receive non-vanishing singular or finite contributions --- after all we are
integrating over zero. On the other hand, in dimensional regularisation there
{\em can}  be non-zero contributions at $O(\epsilon)$ and beyond, simply because
the numerator polynomial ceases to vanish outside $D=4$.

The sorts
of objects which arise in perturbative computations of correlation
functions are traces over $x_{\alpha \dot \alpha}$, and much of this
section will be devoted to understanding 
how to massage such objects into the most useful forms. 
The basic conformal covariants
of trace type are $\text{Tr}(x_{ij} \tilde x_{jk} x_{kl} \ldots \tilde x_{mi})$.
Conformal covariance is guaranteed by the characteristic repetition of points
between neighbouring differences. Due to the index contractions we must have
an even number of entries in the trace. In such a trace we can always use the
manifestly conformal identity
\begin{equation}
x_{13} \tilde x_{32} x_{24} \, = \, - x_{12} \tilde x_{23} x_{34} -
x_{23}^2 \, x_{14} \label{confOrder}
\end{equation}
(here 1,2,3,4 represent any four points) and its conjugate
to put any chain of differences into ascending point ordering. Four-traces
of conformal type do not have a parity-odd sector so that they immediately
reduce to products of squares. For seven points
the longest trace without point repetitions has six entries. There are then
seven such cases according to which point is omitted.

Hence all traces appearing in the calculation reduce to
$\text{Tr}(x_{12} \tilde x_{23} 
 x_{34} \tilde x_{45} x_{50} \tilde x_{01})$ and similar objects.
Point 0 will later be an integration
variable, so that we would like to take it out of the trace by tensor
decomposition in a way that manifestly preserves conformal invariance.
To this end we write the ansatz
\begin{equation}
x_{50} \tilde x_{01} \, = \, \sum_{i = 2,3,4,6} a_i \, y_i \, , \qquad
y_i \, = \, x_{5i} \tilde x_{i1} \label{confTensor}
\end{equation}
because the left hand side is a four-component object and the basis elements
on the right hand side transform in the same way to the left and to the right
as the left hand side does.
Multiplying up by the conjugates of the $y_i$ we obtain four equations
which are indeed invertible. They express the $a_i$
in terms of $x^2_{ij}$ because the conformal traces of length four do not
contain a parity-odd part. \emph{Mathematica} can easily solve the system.
The $a_i$ are found to be polynomials of 27 terms, each composed of six
$x^2_{ij}$ factors, divided by the common denominator
$x^4_{14} x^4_{25} x^4_{36} \, \Delta$,
where $\Delta = (1 - u_1 - u_2 - u_3)^2 - 4 \, u_1 u_2 u_3$.

The numerator polynomials have the appropriate
conformal weights. In particular, there is exactly one $x^2_{i0}$ in every
numerator term. The decomposition can now be substituted into the original
six-trace thereby expressing it in terms of  $x^2_{ij}$ and the trace 1234561.
Since the common denominator is cyclically invariant it is not hard
to derive the decomposition of the other traces involving point 0
from this case.

Finally, the parity-odd part of the trace 1234561 is related to the symbol
$  \sqrt{\Delta}$ (which appears for example in the formulae
(\ref{citePaul}) for the $\rho_6^4$  
component of the R invariants)  simply as\footnote{The term $1-u_1-u_2-u_3$
removes the parity-even part of the trace.}
\begin{equation}\label{eq:18}
\frac{2}{x^2_{14} x^2_{25} x^2_{36}} \,
\text{Tr}(x_{12} \tilde x_{23} x_{34} \tilde x_{45} x_{56} \tilde x_{61}) -
(1-u_1-u_2-u_3) \, = \, \sqrt{\Delta} \, .
\end{equation}
Since the trace 1234561 is essentially unique this is common to
all the R invariants. The sign of the square root in the right-hand side of \p{eq:18} is in fact always positive in the last formula unless $\Delta$ is real and negative. Nevertheless, 
we should keep in mind that the sign of the parity-odd part of the
trace is reversed under the exchange of $x$ for $\tilde x$ implying that
$\sqrt{\Delta}$ is anti-cyclic under shifts.

Finally the six trace can be written in terms of Lorentz objects as 
\begin{eqnarray}
  \sqrt{\Delta} & = & - \frac{2}{x^2_{14} x^2_{25} x^2_{36}} \, \Bigl(
x^2_{26} \, \epsilon(x_{16},x_{36},x_{46},x_{56}) -
x^2_{36} \, \epsilon(x_{16},x_{26},x_{46},x_{56}) + \nonumber \\
&& \phantom{ - \frac{2}{x^2_{14} x^2_{25} x^2_{36}} \, \Bigl( }
x^2_{46} \, \epsilon(x_{16},x_{26},x_{36},x_{56}) \Bigr)\ . \label{nonConfEp}
\end{eqnarray}

\subsection{MHV$_6^{(1)}$  revisited}

In \cite{EKS2} an $x$ space form of the MHV $n$-point 1-loop amplitudes
was derived by evaluating correlators in terms of ${\cal N} = 2$ superfields.
These expressions naturally contained a sum over non-conformal
four-traces like $\text{Tr}(x_{10} \tilde x_{30} x_{40} \tilde x_{50})$.
So as an application of the techniques outlined in the previous
section, let us first try to recast the
integrand of the MHV 6-point 
1-loop amplitude (here taken from the correlator calculation in
\cite{EKS2} in the hexagonal light-like limit - the one-loop correlator
and its integrand are $G_{6;0}^{(1)} = \int G_{7;1}^{(0)}$)
into a convenient form using the aforementioned ideas about
tensor decomposition. We first try to write the MHV integrand in terms
of one-loop integrals of the form
\begin{eqnarray}\label{eq:14}
\widehat A_{6;0}^{(1)}(x_0;x_1,\ldots,x_6) & = & a_1 \, \frac{x^2_{10} x^2_{20} x^2_{35} x^2_{46}}
{x^2_{10} \ldots x^2_{60}} \, + \, \ldots \\ & + & b_1 \frac{
x^2_{10} \, \text{Tr}(x_{23} \tilde x_{34} x_{45} \tilde x_{56} x_{60}
\tilde x_{02})}{x^2_{10} \ldots x^2_{60}} \, + \ldots \nonumber
\end{eqnarray}
where the dots indicate  the 17 other possible terms with a numerator
composed of four
$x^2_{ij}$ (corresponding to 6 one-mass boxes, 6 two-mass hard boxes
and 6  two-mass easy boxes - since the latter can come multiplied by
two different external factors each)
and the 5 other trace terms obtained by cyclic permutations of the
outer points 123456. 
 The existence of a solution follows from the analysis in
\cite{EKS3}. The number coefficients $a_i,b_i$
can be found numerically. The solution is unique up to one
free parameter, which we fix by imposing manifest cyclicity. The
result is then
\begin{eqnarray}
\widehat A_{6;0}^{(1)}(x_0;x_1,\ldots,x_6) & = & 2 \, x^2_{35} x^2_{46} \; g^{\mathrm{1m}}_{12} -
x^2_{25} x^2_{46} \; g^{\mathrm{2mh}}_{13} + (1 - u_3) \, x^2_{25} x^2_{36} \;
g^{\mathrm{2me}}_{14} \nonumber \\
&-& \int \frac{d^4x_0 \,
\text{Tr}(x_{23} \tilde x_{34} x_{45} \tilde x_{56} x_{60} \tilde x_{02})}
{x^2_{20} \ldots x^2_{60}} + (\text{cyclic}) \, .\label{eq:13}
\end{eqnarray}
(The terms with the two-mass easy box double in the cyclic sum.)
Now we can use the tensor decomposition results as outlined around
equation (\ref{confTensor}) 
to  rewrite this in terms of the 1234561 trace (which in turn we write
in terms of $\sqrt{\Delta}$ using~(\ref{eq:18})) and scalar integrals:
\begin{eqnarray}
\widehat A_{6;0}^{(1)}(x_0;x_1,\ldots,x_6) & = &
x^2_{35} x^2_{46} \; g^{\mathrm{1m}}_{12} + \frac{1}{2} (1 - u_3) \, x^2_{25} x^2_{36}
\; g^{\mathrm{2me}}_{14} + (\text{cyclic}) \nonumber \\
&+& \frac{1}{  \sqrt{\Delta}} \; \sum_{i=1}^6 (-1)^i \, \left( \tilde p_i +
\tilde g_{i \, i+2} \right) \label{newMHV} 
\end{eqnarray}
(Note that cycling doubles the two-mass easy terms).
The alternating sign $(-1)^i$ in the sum of integrals in the last line
compensates the anti-cyclicity of $  \sqrt{\Delta}$ (see above) to yield
a cylically invariant result. 

We would like to point the reader's attention to the fact that the
parity-odd part curiously contains the rescaled integrals of the
$\tilde p, \tilde g$ basis, all with coefficients $0,\pm 1$. 

\subsection{NMHV$_6^{(0)}$}

The next amplitude at six-points we wish to obtain from correlation
functions -- as a warm up to the case of interest -- is the
6-point NMHV tree-level amplitude which can be found from the
correlator, $G_{6;1}^{(0)}$ in the hexagonal limit. This is the
simplest non-trivial 
example of the tree-level NMHV correlators considered in the companion
paper \cite{}.  Here we will only consider the component with
$\rho_6$ turned on ie $G_{6;1}^{(0)}|_{\rho_6^4}$ and later will
reconstruct the entire result from this.

As usual we compute the $\cN$=4 correlator
perturbatively using relevant 
${\cal N}=2$ correlators and uplifting to $\cN$=4. Here the $\cN=2$
correlator we need is
\begin{equation}
 \, \lim_{x^2_{i,i+1} \rightarrow 0}
{\langle {O}_1 \, \tilde {O}_2 \, {O}_3 \, \tilde {O}_4 \,
\widehat {O}_5 \, {\cal L}^{{\cal N}=2}_6 \rangle^{(0)}
}
\end{equation}
with
\begin{equation}\label{defop}
{O} \, = \, \mathrm{Tr}(q^2) \, , \quad \tilde {O} \, = \,
\mathrm{Tr}(\tilde q^2) \, , \quad \widehat {O} \, = \, 2 \,
\mathrm{Tr}(q \, \tilde q) \, , \quad{\cal L}_{{\cal N}= 2} \, = \, - \frac{1}{4
\,g^2} \, \mathrm{Tr}(\widehat W^2_{{\cal N} = 2})
\end{equation}
where $q$ is the hypermultiplet and $\widehat W$ the field strength multiplet. See
Appendix~\ref{sec:cal-n=2-superfields} for some facts about ${\cal N} = 2$ superfields. Note that the most
convenient form of the ${\cal N} = 2$ action (as given in the appendix) uses
a field rescaling of the Yang-Mills prepotential $V$ as opposed to standard
conventions. At the linearised level this amounts to
$\widehat W_{{\cal N} = 2, \mathrm{lin}} = g \, W_{{\cal N} = 2, \mathrm{lin}}$.


The correlator calculation proceeds in almost the same way as in the case of the
MHV $n$-point one-loop amplitudes described in \cite{EKS2}. 
 For the sake of
brevity we only point out some differences between the discussion given
there and the new case considered here.

Both the MHV five-point one-loop
amplitude and the NMHV six-point tree level can be found from
this ${\cal N}=2$ correlator. 
In order to obtain the MHV amplitude we put the positions
of the hypermultiplet bilinears at the vertices of a pentagon with light-like
edges. The position of the Lagrangian operator is then the integration
variable of the one-loop MHV integrand. 

In order to obtain the six-point NMHV tree-level amplitude we rather 
place all six operators at the vertices of a hexagon. Due to the different
light-cone limit the range of relevant diagrams becomes slightly smaller and
the cancellation of harmonics with negative charge follows a different
pattern; for instance, the limit selects exactly one ``TT-block'' (c.f.
\cite{EKS2}).
In close parallel to the MHV cases, the light-cone limit is blind to the
actual hypermultiplet projections (i.e. the positioning of ${O},
\tilde {O}, \widehat {O}$) at points 1...5, up to a constant of
proportionality. 
The parity-odd terms sum into $  \sqrt{\Delta}$ via
formula (\ref{nonConfEp}) and the result, after lifting to $\cN$=4
using the techniques in appendix~\ref{sec:cal-n=4-correlators} is simply
\begin{equation}
\lim_{x^2_{i,i+1} \to \, 0} \left. \frac{G^{(0)}_{6;1}}{G^{\text{tree}}_{6;0}} \,
\right\vert_{\rho_6^4} \, = \,  \ 2  
 \left. \Bigl( R_1 + R_3 + R_5 \Bigr )\right\vert_{\rho_6^4} \, 
\, . \label{leadingN2}
\end{equation}
As always at 6-points,  we use the symbol 
$R_1 \,= \, R_{23456}$ (and cyclic) to simplify the notation.
It is crucial to note that since we are only considering the
$\rho_6^4$ component here, we can not immediately reconstruct the full
correlator using superconformal symmetry (which allows us to freely
set  4 $\rho$'s to zero -- a fact we have used extensively throughout
section~\ref{useESS} --  but no more than 4).
Nevertheless we will find that the formula (\ref{leadingN2}) does in
fact 
lift up to the full ${\cal N}=4$ correlator as will be explained in
Section \ref{liftSec} below. 
The result is then in full agreement with our duality conjecture because
$R_1 + R_3 + R_5  \, = \, \widehat A_{6;\text{NMHV}}^{\text{tree}}(x_1,
\ldots,x_6)$.

\subsection{NMHV$_6^{(1)}$}

Let us finally turn to the main aim of this section, the one-loop
correction to the 6-point correlator ${G}_{6;1}^{(1)} $ at order
$\rho^4$ in the light-like limit, which should give the 6-point NMHV 1
loop amplitude according to our duality, and  
which we can obtain as an integral of the 7-point tree-level
correlator $G_{7;2}^{(0)}$ over the position of the additional
Lagrangian at point 0. Again we will first concentrate on a particular
component by switching off $\rho_1, \dots , \rho_5$ and later explain
how to obtain the full result from this. So we wish to compute 
\begin{equation}
 \lim_{x^2_{i,i+1} \rightarrow 0}
{G}_{6;1}^{(1)}(1,\ldots,6)|_{\rho_6^4} \  = \ \int d \mu_0 \, \lim_{x^2_{i,i+1} \rightarrow 0}
G_{7;2}^{(0)}(0;1,\dots,6)|_{\rho_0^4 \rho_6^4} 
\, .
\end{equation}

Again we can't do the perturbative calculation in $\cN$=4 directly,
so instead we compute the relevant $\cN$=2 correlator and reconstruct
$G_{7;2}^{(0)}$. Here the $\cN=2$ correlator is
\begin{equation}
\langle  {\cal L}^{{\cal N}=2}_0 \, {O}_1 \, \tilde {O}_2 \, {O}_3 \,
\tilde {O}_4 \, \widehat {O}_5 \, {\cal L}^{{\cal N}=2}_6 \,
\rangle^{(0)}\ . \label{eq:15}
\end{equation}

As before, the five hypermultiplet bilinears and the Lagrangian at point 6
are at the vertices of the hexagon 1234561 with light-like edges. As in
previous cases, this calculation is simply a different light-cone
limit of a previously studied correlator, 
${G}_{7;2}^{(0)}(x_0,x_{0'};x_1,\ldots,x_5)$  which was used in \cite{EKS2} to
study the five-point two-loop MHV amplitude integrand.

This time however by considering this  new limit one selects a somewhat
different set of graphs. Harmonic analyticity still implies the absence
of the ``TT blocks'', and we can
avoid the systematic use of the cyclic identity for harmonics by the
same identification tricks as in \cite{EKS2}. Since $x_{60}$ is not on the
light-cone, the superconformal reconstruction technique first developed in
\cite{ESS} does not interfere with the light-like limit.
The technique directly yields conformally covariant traces of the type
$x_{16} \tilde x_{62} x_{20} \tilde x_{03} \ldots$
where the points 6 and 0 alternate so that the complete traces must have
length 4,8,12 etc. Putting the points into ascending order (with point 0 beyond
point 6) by the manipulation (\ref{confOrder}), we end up with $x^2_{ij}$ and
the six-traces discussed in Section \ref{traceSec}.

Taking the  light-like limit for the $x^2_{ij}$ terms is
straightforward, but the trace terms are more problematic.
In general off the light cone, there are six independent harmonic
``channels''. But similarly to what we have seen in previous cases on
the light cone most of these channels are subleading, and only the pentagon
one $y_{12}^2 \dots y_{61}^2$ (or in this case the $\cN$=2 analogue remains.)
This
vanishing of all but the pentagon one becomes
manifest upon tensor reducing the traces involving $x_0$. In this channel
we can split the remaining six-trace into its even and odd parts, whereby
the even part reduces to the integrals $p,g$ times some coefficients similar
to those defining the $\tilde p, \tilde g$ basis, while the parity-odd part
still has the terms $x^2_{10} \ldots x^2_{50}  (x^2_{60})^3$ in the common
denominator and thus seems
to diverge. It is possible, though, to take the product of the
vanishing polynomial~(\ref{wow})
 with some smaller expression out of the
numerator in such a way that  $x^4_{60}$ factors out. The odd-part
then reduces to 
the usual 21 scalar integrals, too.

At this point we get the one-loop correlator in a much nicer form and 
which looks like it should simplify further,
 but we have not yet obtained a concise form for it.

The duality conjecture~(\ref{corConj2}) predicts a relation between the
correlator and the square of the amplitude. Expanding out the square
at this order  we thus 
expect a  term of the form (NMHV tree)(MHV one-loop)  in the duality
relation.
Notice that here both factors have a parity-even and a parity-odd
part. We expect the correlator to be related to the NMHV six-point
one-loop amplitude integrand. We know that any  amplitude can be
written (at the level of the {\em integral} in the four-dimensional limit)
as a combination of boxes times coefficients~\cite{Bern:1994zx}. In
this context the amplitude is given as~\cite{annecySuperspace}
\begin{equation}\label{eq:17}
{\widehat \cA}_{6;1}^{(1)}(x_1,\ldots,x_6) \, = \, (R_1 + R_4) \, x^2_{35} x^2_{46} \;
g_{12}^{\mathrm{1m}} + (R_3 + R_6) \, x^2_{25} x^2_{46} \; g_{13}^{\mathrm{2mh}} +
(\text{cyclic}) \, .
\end{equation}
Here it is important that in this equation both sides are the {\em{integrals}} not the
integrands as we shall see. So we now have expressions for both the
correlator and the amplitude in terms of box and pentagon integrals
$p,g$. 
Mellin-Barnes representations for the $p,g$ integrals are straightforward
to derive, so 
with the help of the \emph{MB.m} package \cite{czakon} we have been
able to check 
to satisfactory precision (exact for the singularities, about $0.0(2)$ for
the finite part) that as integrals (where we regularise via standard 
dimensional regularisation for the amplitude and the analogous
regularisation for the correlator and where we have lifted to $\cN=4$)
we have
\begin{eqnarray}
&&{{G}_{6;1}^{(1)} \over G_{6;0}^{(0)}}(x_1,\ldots,x_6)|_{\rho_6^4} \\ 
&& \ =  \ 2 \, \left[ (R_1 + R_3)|_{\rho_6^4} \,
{\widehat \cA}_{6;0}^{(1)}(x_1,\ldots,x_6) + {\widehat  \cA}_{6;1}^{(1)}(x_1,\ldots,x_6)|_{ \rho_6^4}
\right]\ . \nonumber 
\end{eqnarray}
just as predicted by the duality (the first term arises from expanding
out the square of the amplitude to this order).

Now comes the important question of whether this duality can be promoted to an
{\em integrand} identity. The integrand for the correlator as we
define it is simply the correlator
$G_{7;2}^{(0)}$, via the insertion formula~(\ref{eq:3})  $G_{6;1}^{(1)}=\int d\mu G_{7;2}^{(0)}$. 
We do not expect the naive NMHV integrand, $\widehat R_{6;1}^{(1)}$, obtained
from~(\ref{eq:17})  by simply removing  the integrations from the
boxes to lead to an integrand identity.
Instead we have at the
integrand level:
\begin{eqnarray}
&& \int d^4 \rho_0 {{G}_{7;2}^{(0)}(0;1,\ldots,6)|_{\rho_0^4 \rho_6^4} \over G_{6;0}^{(0)}(1,\ldots ,6)} \, = 
\label{n2IntId} \\ && \quad 2  \left[ \, (R_1 + R_3)|_{\rho_6^4} \,
\widehat A_{6;0}^{(1)}(x_0;x_1,\ldots,x_6) \, + \, \widehat R_{6;1}^{(1)}(x_0;x_1,\ldots,x_6
)|_{\rho_6^4} \, + \, r|_{\rho_6^4} \, \right] \nonumber
\end{eqnarray}
where $r$ must vanish upon integration and $\widehat A_{6;1}^{(1)}=\widehat R_{6;1}^{(1)} +r $
becomes the
prediction of our duality for the true NMHV$_6^{(1)}$ integrand.  Note that $r$ has both a
parity even and a parity odd part. 
Suppose we write $  \sqrt{\Delta} = \Delta /  \sqrt{\Delta} $.
The parity-odd part of the last formula then becomes a linear equation
which we can 
easily solve for the parity-odd part of the remainder $r$.
The solution takes a simple form if the integrand identity (\ref{wow})
between the $\tilde p, \tilde g$ integrals is used to eliminate $\tilde p_6$
from the
one-loop MHV amplitude (\ref{newMHV}):
\begin{eqnarray}
r_{\text{odd}}|_{\rho_6^4} & = & \frac{1}{  \sqrt{\Delta}} \; R_{2
  \, \text{even}}|_{\rho_6^4}  \Bigl( 2 \ (\tilde p_2 + \tilde p_5) -
(\tilde g_{12} + \tilde g_{23} + \tilde g_{45} + \tilde g_{56})
\nonumber \\ && \phantom{\frac{1}{  \sqrt{\Delta}} \;
R_{2 \ \text{even}, \ \theta_6^4} \Bigl(}
+ (\tilde g_{24}+ \tilde g_{35}+ \tilde g_{51}+ \tilde g_{62}) - 2 \ \tilde
g_{25} \Bigr) \label{n2FormR} \\ &+&
\frac{1}{  \sqrt{\Delta}} \; R_{3
  \, \text{even}}|_{\rho_6^4} \Bigl( 2 \ (\tilde p_1 + \tilde p_4) -
(\tilde g_{12} + \tilde g_{34} + \tilde g_{45} + \tilde g_{61}) \nonumber \\ &&
\phantom{\frac{1}{  \sqrt{\Delta}} \; R_{3 \ \text{even}, \
\theta_6^4} \Bigl(} + (\tilde g_{13}+\tilde g_{24}+\tilde g_{46}+\tilde g_{51}) -
2 \ \tilde g_{14} \Bigr)\ . \nonumber
\end{eqnarray}
Here $R_{2,3 \ \text{even}}|_{\rho_6^4 }$ refers to the expressions (\ref{citePaul})
with $  \sqrt{\Delta}$ put to zero. Miraculously, if we
upgrade the even part of $R_{2,3}|_{\rho_6^4}$ to the full expressions
including the parity-odd square root terms then  (\ref{n2IntId}) turns
into an \emph{integrand identity} also in the parity-even sector. We
find
\begin{eqnarray}
r|_{\rho_6^4} & = & \frac{1}{  \sqrt{\Delta}} \; R_{2
 }|_{\rho_6^4}  \Bigl( 2 \ (\tilde p_2 + \tilde p_5) -
(\tilde g_{12} + \tilde g_{23} + \tilde g_{45} + \tilde g_{56})
\nonumber \\ && \phantom{\frac{1}{  \sqrt{\Delta}} \;
R_{2 \ \text{even}, \ \theta_6^4} \Bigl(}
+ (\tilde g_{24}+ \tilde g_{35}+ \tilde g_{51}+ \tilde g_{62}) - 2 \ \tilde
g_{25} \Bigr) \label{n2FormR2} \\ &+&
\frac{1}{  \sqrt{\Delta}} \; R_{3
 }|_{\rho_6^4} \Bigl( 2 \ (\tilde p_1 + \tilde p_4) -
(\tilde g_{12} + \tilde g_{34} + \tilde g_{45} + \tilde g_{61}) \nonumber \\ &&
\phantom{\frac{1}{  \sqrt{\Delta}} \; R_{3 \ \text{even}, \
\theta_6^4} \Bigl(} + (\tilde g_{13}+\tilde g_{24}+\tilde g_{46}+\tilde g_{51}) -
2 \ \tilde g_{14} \Bigr)\ . \nonumber
\end{eqnarray}
\emph{Mathematica}'s inbuilt factorisation algorithm can show this
algebraically; alternatively one can
substitute numbers. The \emph{MB.m} package shows to good precision that the
two sums of integrals in (\ref{n2FormR2}) separately integrate to zero.
We remark that both sums can be made to contain only coefficients $0,\pm 1$
by adding in the integrand identity (\ref{wow}).

In summary then we have a clear prediction for the NMHV$_6^{(1)}$
integrand albeit only in the $\rho_6$ sector. We now turn to the
question of obtaining the full integrand in all sectors, ie
obtaining the full $\rho$ dependence of $r$.

\subsection{Reconstructing the full supercorrelator/ NMHV amplitude} \label{liftSec}

We now wish to try to reconstruct the full one-loop correlator
$G_{6;1}^{(1)}$ with its full $\rho$ dependence in the hexagon limit
from the results of
the previous section where the $\rho_6$ projection was derived from
$\cN$=2 correlators. In fact we eventually discuss the integrand, which is
simply the tree correlator $G_{7;2}^{(0)}$.

 Our conjecture is that the hexagon light-cone limit will
reproduce the  NMHV six-point one-loop 
amplitude and related terms and indeed we have already seen this at
the level of the integral and for the $\rho_6$ projection.

In order to perform this reconstruction, we assert that the
correlation function in the hexagon  limit must be a linear
combination of NMHV 6-point $R$ invariants, $R_1 \ldots R_6$ times
pseudoconformal integrals.  This is simply because the correlator is superconformally invariant, and in the hexagon limit $R_i$ are all
the superconformal nilpotent invariants at this order. (The analysis
is identical to the corresponding amplitude analysis performed
in~\cite{annecySuperspace} translated to analytic
superspace.) It could be that there are more nilpotent invariants off the
light-like limit, but we will not look into this issue here.

Since we know that $R_1+R_3+R_5 \, = \, R_2+R_4+R_6$, only $R_1 \dots
R_5$ are independent and  we therefore have in principle the
following expansion for $G_{6;1}$ (indeed the same expansion is valid
at any loop order so for the moment we do not specify the loop order) 
\begin{eqnarray}
G_{6;1} & = & R_1 \left( y_{12}^2 y_{23}^2 y_{34}^2 y_{45}^2 y_{56}^2 y_{61}^2
\, f_{11}(x) + y_{12}^4 y_{34}^4 y_{56}^4 \, f_{12}(x) + \ldots \right) \nonumber \\
& + & R_2 \left( y_{12}^2 y_{23}^2 y_{34}^2 y_{45}^2 y_{56}^2 y_{61}^2
\, f_{21}(x) + y_{12}^4 y_{34}^4 y_{56}^4 \, f_{22}(x) + \ldots \right) \label{ans650}
\\ & + & \ldots \nonumber \\ & + & R_5  \left( y_{12}^2 y_{23}^2 y_{34}^2 y_{45}^2
y_{56}^2 y_{61}^2 \, f_{51}(x) + y_{12}^4 y_{34}^4 y_{56}^4 \, f_{52}(x) +
\ldots \right) \nonumber
\end{eqnarray}
The presence of these $y$ terms is a straightforward consequence of
the $SU(4)$ R symmetry. One needs to write down any monomial in
$y_{ij}^2$ 
for which all indices 1,2, ..., 6  appear precisely twice.  There are
a total of 130 distinct non-vanishing harmonic structures with 
charge 2 at every point, of which we have displayed the hexagon one (which will
finally be the only one to survive) and --- for illustration purposes ---
one other term. 

However considerations of this formula projected on the $\rho_6^4$
component vastly reduce the number of independent functions.
 The $\rho_6^4$
components of the R invariants, which are displayed in equation
(\ref{citePaul}) have a universal form
\begin{equation}
R_i|_{\rho_6^4} \, = \, R_{i6}(u_1,u_2,u_3,  \sqrt{\Delta}) \,
\frac{\rho_6^4 \, x^2_{24}}{x^2_{26} \, x^2_{46}} \, \frac{y^2_{15}}{y^2_{16} \,
y^2_{56}} \, , \qquad R_{56} \, = \, R_{66} \, = \, 0 \, ,
\end{equation}
where this equation defines the functions $R_{ij}$ for all  $i,j$ by
cyclic rotations. The $\rho_i^4$ components of the $R$
invariants introduce singularities in $y^2$ into \p{ans650}. Since
the whole correlator is supposed to contain only finite
dimensional representations of the internal symmetry group we
require the absence of such poles; in other words we demand
``harmonic analyticity''. This is a rather strong constraint which
puts most of the unknown functions in \p{ans650} to zero.

Next, we also know that in the light-like limit the $y$ dependence
of $G_{6;1}|_{\rho_6^4}$ is just given by the pentagon harmonic
structure $y_{12}^2y_{23}^2y_{34}^2y_{45}^2y_{51}^2$ (this is
admittedly currently an observation on the relevant $\cN = 2$
correlators rather than a proven feature -- it is certainly true
in the tree and one-loop cases we are interested in as seen in the
previous two sections and it seems likely to be true in general at
any loop order).

But then taking the $\rho_6^4$ projection of~\eqref{ans650} together
with these facts we see that the only allowed harmonic
structure 
in~\eqref{ans650} is the the hexagon $y$-structure. Namely we have
\begin{equation}
\lim_{x_{i i+1}^2 \rightarrow 0 } { G}_{6;1} \, = \, y_{12}^2 y_{23}^2 y_{34}^2 y_{45}^2 y_{56}^2 y_{61}^2 \, \sum_{i=1}^5
R_i \, f_i(x) \, . \label{finallyOnlyRTimesF}
\end{equation}
The $f_i(x)$ are obviously subject to cyclic invariance.

Finally we can determine the 5 functions $f_i(x)$ by using the
perturbative computations of the previous two sections, 
where we computed 
the $\rho_6$ component of this correlator at tree-level and one-loop, 
$G_{6;1}^{(l)}|_{\rho_6^4}$, for $l=0,1$ together with its cyclic shifts.

In particular from here we can reconstruct the full function $r$
(recall $r$ is the
difference between our prediction for the full amplitude integrand and the
naive integrand $\widehat R_{6;1}^{(1)}$ involving box integrands  only.)

We certainly can not simply take the result (\ref{n2FormR}) for $r$
and remove the $|_{\rho_6^4}$ since for one thing the result  is not
cyclically invariant.
 Similarly to the result for ${G}_{6;1}$ (\ref{finallyOnlyRTimesF}) we
 have
 that $r=\sum_i R_i g_i(x)$ 
 for $g_i(x)$ to be determined. But the results of the last
 section, give us (see (\ref{n2FormR2})
\begin{equation}
\sum_{i=1}^5 R_{i6} \, g_{i}(x)  \, = \, r|_{\rho_6^4} \, = \, \text{known}
\end{equation}
and the five cyclic shifts of this equation, which constitutes
a system of six equations for five unknowns $g_i(x)$. We have sought a solution
for five of the equations, separately for the even and the odd part of the 
$\rho_i^4$ components of the R invariants, using their explicit dependence
on $u_1,u_2,u_3$. The sixth equation is satisfied
in both cases thanks to the identity (\ref{wow}) between the 21 scalar
integrals. The solutions  as computed from the even
and the odd part look different at the first glimpse, but upon eliminating
the $\tilde p_5$ integral by equation (\ref{wow}) one finds  the same
solution in both cases. Curiously, the
final form of $r$ is 
\begin{eqnarray}
r & = & \frac{1}{  \sqrt{\Delta}} \; (R_1 - R_4) \,
\Bigl( 2 \ (\tilde p_1 + \tilde p_4) -
(\tilde g_{12} + \tilde g_{34} + \tilde g_{45} + \tilde g_{61})
\nonumber \\ && \phantom{\frac{1}{  \sqrt{\Delta}} \; (R_1 - R_4) \, \Bigl(}
 + (\tilde g_{13}+\tilde g_{24}+\tilde g_{46}+\tilde g_{51}) -
2 \ \tilde g_{14} \Bigr) \label{yes} \\ &+&
\frac{1}{  \sqrt{\Delta}} \; (R_2 - R_5) \, \Bigl( 2 \ (\tilde p_3 + 
\tilde p_6) -
(\tilde g_{23} + \tilde g_{34} + \tilde g_{56} + \tilde g_{61}) \nonumber \\ &&
\phantom{\frac{1}{  \sqrt{\Delta}} \; (R_2 - R_5) \, \Bigl(} +
(\tilde g_{13}+\tilde g_{35}+\tilde g_{46}+\tilde g_{62}) -
2 \ \tilde g_{36} 
 \Bigr) \nonumber \, .
\end{eqnarray}
As in formula (\ref{n2FormR}) the two sums of scalar integrals 
can be made to contain only coefficients $0,\pm1$ by the integrand
identity (\ref{wow}).
The complete expression (\ref{yes}) is cyclically invariant: after a shift
(respecting the anti-cyclicity of $  \sqrt{\Delta}$) one may replace
$(R_3 - R_6) \, = - (R_1 - R_4) + (R_2 - R_5)$ and finally use (\ref{wow})
to restore the original form.

While the $\tilde p, \tilde g$ integrals in the integrand identity (\ref{wow}),
the linear combination in the parity-odd part of the MHV amplitude
(\ref{newMHV}), and the two sums of integrals in (\ref{yes}) add up to zero
at $O(1/\epsilon^2), \, O(1/\epsilon), \, O(1)$ in all four cases,
the \emph{MB.m} package clearly indicates otherwise\footnote{The
cancellation of the singularities works to very high
precision, while the finite parts typically yield 0.0(2). At $O(\epsilon)$
we consistently found non-vanishing contributions to all four sums;
for some sample points even rather large numbers. The first three
digits were always significant according to the error estimates
supplied by the package.} at $O(\epsilon)$. Since we have employed
(\ref{wow}) in many places in our analysis, our result (\ref{yes}) is not
necessarily valid with respect to  $O(\epsilon)$ corrections in dimensional
regularisation or related schemes.

Summing up then we have that the integrand of
$G_{6;1}^{(1)}$, namely $G_{7;2}^{(0)}$ is given by:
\begin{align}
 \lim_{x_{i \, i+1}^2 \rightarrow 0}   \int d^4 \rho_0 &  \frac{G_{7;2}^{(0)}(0;1,\ldots,6)}
{G_{6;0}^{(0)}(1,\ldots,6)} \, \bigg\vert_{\rho_0^4}  \label{n4IntId}
\\ \notag
& 
\, = \, 2 \,  \left( \widehat A_{6;1}^{(0)}(1,\ldots,6) \,
\widehat A_{6;0}^{(1)}(0;1,\ldots,6) \, + \, 
\widehat R_{6;1}^{(1)}(0;1,\ldots,6) \, + \, r \right)  
\end{align}
as an integrand identity in four dimensions.

\subsection{Match with  the amplitude integrand proposal} \label{numSec}

We now wish to  compare our predicted amplitude integrand to the
corresponding expression 
in \cite{nima2} which is written in terms of momentum
twistors 
\begin{equation}
Z_i=(\lambda_i,\mu_i)\,,\qquad 
 \mu_{i\da}=\lambda_i^\a (x_i)_{\a\da}\,, \qquad \tilde x_i^{\da\a} = \frac{\lambda_i^\a \mu_{i-1}^\da-\lambda_{i-1}^\a \mu_{i}^\da}{\vev{i-1\,i}}\,.
\end{equation}
A local form of the output of generalised BCFW rules \cite{BCFW} for the integrand of
all one-loop NMHV amplitudes was given in \cite{nima1}. Specialising
to the six-point case this gives
\begin{eqnarray}
\widehat A_{6;1}^{(1)}(x_0;x_1,\ldots,x_6) & = & \sum_{i=1}^6 \, (J_{i+1,i+2,i+4} + J_{i+3,i+4,i+1} +
J_{i+5,i+1,i+3}) \, R_i  \nonumber \\
& + & \sum_{i=1}^6 \, J_{i,i+2} \, R_{i+1} \, +  \, J_{i,i+1} (R_1
+ R_3 + R_5) \, , \label{newNima}
\end{eqnarray}
where
\begin{eqnarray}\notag
J_{i,j} & = & \frac{\la A B \, (i-1 \; i \; i+1) \cap (j-1 \, j \, j+1) \ra \la
X i j \ra}
{\la A B X \ra \la A B \, i-1 \, i \ra \la A B \, i \; i+1 \ra \la A B \, j-1
\, j \ra \la A B \, j \, j+1 \ra} \, , \\
J_{i,j,k} & = & \frac{\la A B (i-1 \; i \; i+1) \cap \Theta_{ijk} \ra}
{\la A B X \ra \la A B \, i-1 \, i \ra \la A B \, i \; i+1\ra \la A B \, j \, 
j+1 \ra \la A B \, k \, k+1 \ra} \, .
\end{eqnarray}
In these formulae $(A B)$ define the integration point and
$X \, = \, (C D)$ is an arbitrary bispinor of which the
six-point integrand (\ref{newNima}) is in fact independent. The $\cap$
symbols mean intersections in twistor geometry following the rules
\begin{eqnarray}\notag
\la A B \, (i \, j \, k) \cap (l \, m \, n) \ra & = &
\la A \, i \, j \, k \ra \la B \, l \, m \, n \ra -
 \la A \, l \, m \, n \ra \la B \, i \, j \, k \ra \, , \\
(i \, j \, k) \cap X & = & D \, \la C \, i \, j \, k \ra -
C \, \la D \, i \, j \, k \ra
\end{eqnarray}
and the surface $\Theta_{ijk}$ is defined as
\begin{equation}
\Theta_{ijk} \, = \, \frac{1}{2} \left[ (j \, j+1 \, (i \; k \, k+1) \cap X) - 
(k \, k+1 \, (i \; j \, j+1) \cap X) \right] \, .
\end{equation}
Through the chain of back-substitutions the NMHV integrand (\ref{newNima}) is
reduced to twistor four-brackets as defined in (\ref{detForm}) and $R$
invariants.

It remains to compare formula (\ref{newNima}) to our prediction for
this integrand from the 
six-point correlator, namely $\widehat R_{6;1}^{(1)} +r$ (see~\p{yes}). 
So we wish to  verify
\begin{equation}
\widehat R_{6;1}^{(1)} \, + \, r \, = \,
\widehat A_{6;1}^{(1)} \, , \label{glorious}
\end{equation}
where the left-hand side is our prediction and
the right hand side is the local twistor integrand (\ref{newNima}).
Since the $\rho_1 \ldots \rho_6$ structure of the invariants is rigid and
${\cal N}=4$ correlator and amplitude are cyclically invariant we will not
do this for more than one Grassmann component; given the discussion in the
preceding sections the obvious choice for us will be $\rho_6^4$.
If random complex integers are chosen for the components of
$Z_1, \ldots, Z_6, A, B, C, D$ (we limited the range to $[-100,100]$ for
real and imaginary parts) the evaluation of either side of (\ref{glorious})
stays in the rational numbers which \emph{Mathematica} can treat exactly;
any disagreement would be noticed. The evaluation of the twistor integrand
simply uses
the determinant form of the $\la i j k l \ra$ four-bracket, while
our $x$-space integrand can be calculated by matrix multiplication
after gaining the $x_i$ from the twistors by \p{makeX}.

We have successfully run this check for hundreds of sample points confirming
that the correspondence between correlation functions (as calculated by
Lagrangian insertions in the ${\cal N} = 2$ formalism) and amplitudes
holds at the loop integrand level for NMHV cases, too.

Last, according to \cite{nima2}
\begin{eqnarray}
\cA_{5,1}^{(1)} & = & R_{135} \sum_{i=1}^5 \left( J_{i,i+1,i+3} +
J_{i,i+1} \right) \, , \\ \cA_{5,0}^{(1)} & = & \sum_{i<j} J_{i,j}
\end{eqnarray}
from which we may check numerically that
\begin{equation}
\cA_{5,1}^{(1)} + R_{135} \cA_{5,0}^{(1)} \, = \, R_{135} \left( 4
\pi^2 \, x_{13}^2 x_{24}^2 \, g(1,2,3,4) + \text{cyclic} \right)
\end{equation}
as stated in the main text.

\section{Conclusions}

We have illustrated with several examples that the tree-level $n+m$-point function of
${\cal N} = 4$ stress tensor multiplets generates all N$^k$MHV $(n+k)$-point
$(m-k)$-loop amplitude integrands  for $0 \leq k \leq m$ in the appropriate light-like
limits. This extends the correlator/MHV amplitude duality discussed in
\cite{EKS2,EKS3} to the {\em entire} super-amplitude in planar $\cN=4$ SYM. The feature that a given correlator can generate a variety
of amplitudes has a close parallel in the super Wilson
loop proposal of \cite{marksAndSparks,simon}.

We conjecture that the correlator/amplitude duality generally holds at
the level of the \emph{integrand}. As a highly non-trivial test we have
used the new correspondence to construct the integrand of the
NMHV six-point one-loop amplitude and confirmed its exact equality with the
corresponding prediction of the BCFW recursion rules for the all-loops
integrand \cite{nima1} in local form \cite{nima2}. To compare
the two expressions we have substituted random generated complex rational
numbers, which \emph{Mathematica} can manipulate without numerical
approximations.

The correlator computation relevant to the latter check was done entirely
in the traditional $x$ space variables, using
only conformally covariant manipulations for the reduction and tensor
decomposition of traces over $x^{\alpha \dot \alpha}$. The final $x$ space
formulae for the parity-odd sector of the
integrand are strikingly simple; if a certain basis is used for the scalar
integrals we find only coefficients $0,\pm 1$. We hope that this circle of
ideas will be useful in other applications, too.

\section*{Note added}

When this paper was ready for submission, two other publications on the duality between correlators and Wilson loops appeared \cite{Bianchi:2011rn}, \cite{Adamo:2011dq}. The former treats the duality in three dimensions, the latter proposes a twistor superspace version of it.

\section*{Acknowledgements}

ES is grateful to  Sergio Ferrara, Paul Howe  and Raymond Stora
and PH is grateful to Valya Khoze and Claude Duhr for a
number of enlightening discussions. GK and ES acknowledge discussions
with David Skinner and Simon Caron-Huot. BE and PH acknowledge support
by STFC under the rolling grant ST/G000433/1. BE acknowledges hospitality
at the ITP Leipzig during the final stages of this work.

\appendix

\section{Partial non-renormalisation of the four-point correlator $G_{4;0}$}
\label{sec:part-non-renorm}
In the minimal on-shell $\cN = 4$ analytic superspace formalism \cite{paulN4}
there is the equivalent { of the $x$-space conformal inversion} acting on the internal $y$
variables.
The stress-energy tensor multiplet has weight $(-2)$ { (to be identified with the $SU(4)$ harmonic charge $(+2)$ used in \cite{twin}) under this transformation. This defines it as the highest-weight state of the irrep ${\bf 20'}$ of $SU(4)$. At the same time, it has conformal weight two.}
At $\rho=\bar \rho = 0$ the full structure of the four-point correlator is
\begin{align}\label{4ptdet}
G_{4;0}(x_1, \ldots, x_4)
&= \frac{y^2_{12}y^2_{23}y^2_{34}y^2_{41}}{x^2_{12}x^2_{23}x^2_{34} x^2_{41}}F_{1} +\frac{ y^2_{12}y^2_{13}y^2_{24}y^2_{34}}{x^2_{12}x^2_{13}x^2_{24} x^2_{34}}F_{2} +\frac{y^2_{13}y^2_{14}y^2_{23}y^2_{24}}{x^2_{13}x^2_{14}x^2_{23} x^2_{24}}F_{3}  \nt
&+ \frac{y^4_{12} y^4_{34}}{x^4_{12}x^4_{34}}F_{4}  + \frac{y^4_{13} y^4_{24}}{x^4_{13}x^4_{24}}F_{5} + \frac{y^4_{14} y^4_{23}}{x^4_{14}x^4_{23}}F_{6}   \,,
\end{align}
where $F_i(s,t;a)$ (with $i=1,\ldots,6$) are functions of the two independent
conformal cross-ratios
\begin{align}\label{}
s=\frac{x^2_{12}x^2_{34}}{x^2_{13}x^2_{24}}\,, \qquad
t=\frac{x^2_{14}x^2_{23}}{x^2_{13}x^2_{24}}\,, 
\end{align}
as well as of the `t Hooft coupling $a$. 
The six terms in (\ref{4ptdet}) correspond to the six $SU(4)$ irreps in
the tensor product ${\bf 20'}\times{\bf 20'} = {\bf 1}+{\bf15}+{\bf 20}+
{\bf 84}+{\bf 105}+{\bf 175}$. Each of them consists of a propagator structure
and a conformally invariant function. The propagator structures are
obtained by connecting the four points with free propagators \p{treeNGon} in
all possible ways (Wick contractions). They have the required conformal weight
two and internal charge two at each point. { At tree level} the first three terms are described
by connected, the other three terms by disconnected graphs. The six
coefficient functions $F_i(s,t;a)= C_i + a F^{(1)}_i(s,t) + a^2 F^{(2)}_i (s,t)+
\ldots$ comprise tree-level constants and functions from loop corrections.

A very important property of the four-point correlator is that the six loop
correction functions are not independent: they are all proportional to a single
function of $s,t$. The loop correction to the correlator, i.e. the part
excluding the tree-level contribution, takes the following factorised form
\begin{align}\label{deco}
G_{4;0}  - G^{(0)}_{4;0} \, = \,  I(x_1,\ldots,x_4, y_1, \ldots, y_4) \, F(s,t;a)\,,
\end{align}
with the rational prefactor
\begin{align}\label{panon}
 I &= \frac{y^2_{12}y^2_{23}y^2_{34}y^2_{41}}{x^2_{12}x^2_{23}x^2_{34} x^2_{41}}(1-s-t)
+\frac{ y^2_{12}y^2_{13}y^2_{24}y^2_{34}}{x^2_{12}x^2_{13}x^2_{24} x^2_{34}}(t-s-1)  \nt
& +\frac{y^2_{13}y^2_{14}y^2_{23}y^2_{24}}{x^2_{13}x^2_{14}x^2_{23} x^2_{24}}(s-t-1)+
\frac{y^4_{12} y^4_{34}}{x^4_{12}x^4_{34}}s  + \frac{y^4_{13} y^4_{24}}{x^4_{13}
x^4_{24}} + \frac{y^4_{14} y^4_{23}}{x^4_{14}x^4_{23}}t\,. 
\end{align}
We stress that this result is valid to all orders in the coupling. This
remarkable fact, known as ``partial non-renormalisation" \cite{partialNonRen},
can be explained in two equivalent ways:

One explanation is given by the superconformal Ward identities which the
correlator has to satisfy. Apart from the simple fact that the odd variables
of the supercorrelator $\vev{\cT_1 \ldots  \cT_4}$ can be gauged away by $Q$
and $\bar S$ transformations, the various components in its Grassmann expansion
satisfy differential constraints following from the full superconformal
algebra with generators $Q, \bar Q, S$ and $\bar S$. Their general solution
\cite{partialNonRen} involves an arbitrary two-variable function exactly as in \p{deco},
\p{panon}, and in addition a single variable function not shown in \p{deco}.
The detailed analysis of the conformal partial wave expansion of the latter
shows that it can only contribute at tree level, where it takes a fixed form
\cite{Dolan:2001tt,Dolan:2004mu}. The loop corrections, i.e., the derivatives of the correlator with 
respect to the coupling, always take the form \p{deco}, \p{panon}. One can also see this directly in $\cal N$=4 analytic superspace in which the correlator becomes an expansion in $SL(2|2)$ invariants (ie characters or Schur polynomials labelled by $SL(2|2)$ representations). The renormalised two variable function comes directly from generic long (or typical) representations of $SL(2|2)$ whereas the protected one variable functions arise from short representations~\cite{Heslop:2002hp}.

The alternative explanation \cite{ESS} makes use of a Lagrangian insertion:
In formula \p{insertIt} we have stated that the $O(a^l)$ correction to
$G_n$ can be calculated using $l$ insertions. If the integrand on the
right-hand side is restricted to Born level this yields in fact exactly
the $O(a^l)$ part of $G_n$, else all corrections at $O(a^m) , \, m \geq l$
are reproduced.
The one-insertion scenario implies that the loop corrections to the
four-point correlator $G_{4;0}$ are obtained from the nilpotent part $G_{5;1}$ of
the five-point one. The latter is heavily restricted by $\cN=4$ conformal
supersymmetry and has the following general form
\begin{align} \label{}
G_{5;1} =  P(x_1,\ldots, x_5; \rho_1, \ldots, \rho_5; y_1, \ldots, y_5) \,
f(x_1, \ldots,x_5)
\end{align}
at $\bar \rho_i  = 0$. Here $P(x, \rho, y) $ is a very specific nilpotent
rational function of the space-time, odd and harmonic variables of Grassmann
degree four, carrying the necessary conformal weights and internal charges at 
all five points. The only remaining freedom is in the arbitrary function
$f(x)$. Further, the coefficient of the $(\rho_5)^4$ component of $P$ turns out
not to depend on $x_5$ and $y_5$,
\begin{align} \label{4.7}
P_{\rho_1=\ldots=\rho_4=0} =  I(x_1,\ldots,x_4, y_1, \ldots, y_4)\  \rho_5^4\,,
\end{align}
with the same prefactor $I$ as in \p{panon}. 
Integrating out the insertion point we find the
factorised form \p{deco} of the loop corrections, where the arbitrary
two-variable function is given by $F(s,t) = \int d^4 x_5 \, f(x_1, \ldots,x_5)$. 

Now, the practical question is how to compute the loop corrections. The
absence of an off-shell formulation of $\cN=4$ SYM makes Feynman graph
calculations with manifest $\cN=4$ supersymmetry impossible. We have to
resort to component calculations (with no manifest supersymmetry) or to
formulations in terms of $\cN=1$ or $\cN=2$ superfields.  The $\cN=2$
formulation has the advantage that it reproduces the phenomenon of partial
renormalisation, for exactly the same reason as in the $\cN=4$ case.
Here we just give the results of the one- and two-loop computations
\cite{me1,simp,schalm,ESS,Bianchi}:
\begin{align}
F&= \frac{2 N^2_c}{(4 \pi^2)^4} \left[ \frac14 a \, F^{(1)} + \frac1{16} a^2 F^{(2)} + O(a^3) \right]
\, , 
\notag
\\
F^{(1)} &=  \, x_{13}^2 x_{24}^2 \, g(1,2,3,4) \label{4.10'} \, , \\
F^{(2)} &=  \, x_{13}^2 x_{24}^2 \, \biggl[ \frac12(x_{12}^2x_{34}^2+x_{13}^2x_{24}^2+
x_{14}^2x_{23}^2) (g(1,2,3,4))^2\notag
\\[2mm] \notag
& \phantom{\, x_{13}^2 x_{24}^2 \, \biggl[} \quad +
x_{12}^2 h(1,2,3;1,2,4) +  x_{23}^2 h(1,2,3;2,3,4) + 
x_{34}^2 h(1,3,4;2,3,4) \nt
& \phantom{\, x_{13}^2 x_{24}^2 \, \biggl[} \quad +  x_{41}^2 h(1,2,4;1,3,4)
+ x_{13}^2 h(1,2,3;1,3,4) + x_{24}^2 h(1,2,4;2,3,4) \, \biggr] \, \label{4.10}
\end{align} 
where the off-shell one- and two-loop integrals are defined by
\begin{align}  \label{h1}
g(1,2,3,4) &=\frac{1}{4 \pi^2} \int\frac{d^4 x_0}{x_{10}^2x_{20}^2x_{30}^2x_{40}^2}\,,
\\  \label{h12}
h(1,2,3;1,2,4) &= \frac{1}{(4 \pi^2)^2} \int \frac{d^4 x_0 d^4 x_{0'}}{
(x_{10}^2x_{20}^2 x_{30}^2 )x_{0{0'}}^2( x_{1{0'}}^2x_{2{0'}}^2x_{4{0'}}^2)} \,
\end{align}
in four dimensions.

Assembling this altogether gives the formulae quoted
in~(\ref{eq:6}-\ref{eq:9}). In particular the two-loop and one-loop
squared pieces of $F^{(2)}$ reassemble into the suggestive form in~(\ref{eq:9}).

\section{${\cal N}=2$ superfields}
\label{sec:cal-n=2-superfields}
Real ${\cal N} = 2$ Minkowski space has the coordinates $x^{\alpha\dot \alpha},
\theta^{i \alpha},\bar \theta^{i \dot \alpha}$ where $i \in \{1,2\}$. The internal
$SU(2)$ index $i$ can be raised and lowered by $\epsilon^{ij}, \epsilon_{ij}$ like
the Lorentz indices $\alpha, \dot \alpha$.

\emph{Harmonic superspace} \cite{GIKOS} has an additional 
coordinate $u = (u^+,u^-) \in SU(2)/U(1)$. { Instead of choosing a coordinate
representative of the coset one uses the entire matrix $u \in SU(2)$. This helps to preserve the manifest $SU(2)$ and to keep track of the local $U(1)$ charge. (In contrast, in the case of
${\cal N} = 4$ described below we
prefer to work with coordinates on the harmonic coset.) An $SU(2)$ invariant combination of harmonics at two different points in harmonic space is given by}
\begin{equation}
(12) \, = \, u^{+i}|_1 \epsilon_{ij} u^{+j}|_2 \, .
\end{equation}

\emph{Analytic superspace} has the coordinates $\{x, \; \theta^+,\bar
\theta^+, u \}$ where
\begin{equation}
\theta^+ \, = \, \theta^i u^+_i \, , \qquad \bar \theta^+ \, = \,
\bar \theta^i u^+_i
\end{equation}
thus involving only one (plus projected) half of the odd coordinates. The
${\cal N}=2$ matter multiplet (the hypermultiplet $q^+$) and the
Yang-Mills multiplet (incorporated in the gauge prepotential $V^{++}$) can both be realised
as unconstrained quantum fields on analytic superspace \cite{GIKOS}:
\begin{eqnarray}
q^+(x_\cA , \, \theta^+,\bar \theta^+,u) \, , \quad
V^{++}(x_\cA , \, \theta^+,\bar \theta^+,u) \, , \qquad
x_\cA  = x - 4 \, i \, \theta^{(i} \bar\theta^{j)} \, u^+_i u^-_j \, .
\end{eqnarray}
The field content of the ${\cal N}=4$ super-Yang-Mills theory is equivalent
to the physical fields of the two ${\cal N} = 2$ multiplets put together.
The ${\cal N}=4$ action is obtained when both
fields transform in the adjoint representation of the gauge group:
\begin{equation}
 S_{\mathrm{{\cal N}=4 \; SYM}} \, = \, S_{\mathrm{HM/SYM}} + S_{\mathrm{{\cal N}=2 \; SYM}}
\end{equation}
with
\begin{eqnarray}
  S_{\mathrm{HM/SYM}}
    & =  & - 2 \int
dud^4x_\cA d^2\theta^+d^2\bar\theta^+\ {\rm Tr} \left( \tilde q^{+}D^{++}q^+ +
i \, \sqrt{2} \, \tilde q^{+} [V^{++},q^+] \right) \\
  S_{\mathrm{{\cal N}=2 \; SYM}}& = &
- {1\over  4 g^2}\int d^4x_L d^4\theta\;  {\rm Tr}(\widehat W_{{\cal N} = 2}^2) \, ,
\qquad x^{\alpha\dot\alpha}_L = x^{\alpha\dot\alpha} - 2i\theta^{i\alpha} \bar
\theta^{\dot \alpha}_i \label{n2SYM} \\
\widehat W_{{\cal N} = 2} & = & {i\over 4} u^+_iu^+_j \bar D^i_{\dot\alpha}\bar D^{j\dot\alpha}
\sum^\infty_{r=1} \int du_1\ldots du_r\; {(-i \sqrt{2})^{r} V^{++}(u_1)
\ldots V^{++}(u_r) \over (u^+u^+_1)(u^+_1u^+_2) \ldots (u^+_ru^+)}\; .
\nonumber
\end{eqnarray}
In the definition of the field strength $\widehat W_{{\cal N} = 2}$ the auxiliary
harmonics $u_1, \ldots,u_r$ are integrated out. It is less obvious --- 
but nonetheless true --- that the field strength is also independent of
the non-integrated harmonic variable $u$. $\widehat W_{{\cal N} = 2}$ is in fact a
chiral field depending on $x_L,\theta^i$. 
Notice that in our convention
$V^{++}$ has been rescaled with the gauge coupling, $V \to
\sqrt{2} \, g^{-1} V$ w.r.t. the definitions in \cite{GIKOS}, which
has the effect that the coupling is present only in front of the
SYM action \p{n2SYM} and that the physical scalar in $V^{++}$
acquires a propagator with standard normalisation.

In this article we draw upon the Feynman graph calculations of
\cite{ESS,EKS2,EKS3} where the necessary Feynman rules and methods of
calculation are explained in detail. We are interested in correlation functions
of the operators
\begin{equation}
O \, = \, \mathrm{Tr}(q^+ q^+) \, , \qquad \tilde O \, = \,
\mathrm{Tr}(\tilde q^+ \tilde q^+) \, , \qquad \widehat O \, = \, 2 \,
\mathrm{Tr}(\tilde q^+ q^+) \, , \nonumber
\end{equation}
like e.g.
\begin{eqnarray}
{\cal G}_{n;0}(x_1,\ldots,x_n) & = & \la O(x_1) \, \tilde O(x_2) \, O(x_3)
\ldots \tilde O(x_n) \ra \\
& = & \int {\cal D}\Phi\ e^{S_{{\cal N} = 4 \; \mathrm{SYM}}} \
\tilde O(x_1) \, O(x_2) \, \tilde O(x_3) \ldots O(x_n)
\, . \notag
\end{eqnarray}
Differentiation of the path integral with respect to  the coupling constant
yields the identity
\begin{eqnarray}
&& g^2 \frac{d}{d g^2 } \, {\cal G}_{n;0}(x_1,\ldots,x_n) \label{insertOneLoop} \\
&& \; = \, \frac{1}{g^2} \int d^4x_0 d^4\theta_0 \int {\cal D}\Phi\
e^{S
} \
O(x_1) \, \tilde O(x_2) \, O(x_3) \ldots \tilde O(x_n) \, \frac{1}{4 g^2} 
\tr(\widehat W_{{\cal N} = 2}^2) \nonumber \\
&& \; = \, \frac{1}{g^2} \int d^4x_0 d^4\theta_0 
\ {\cal G}_{n+1;1}(x_0;x_1,\ldots,x_6) \, .
\nonumber
\end{eqnarray}
Equations~(\ref{insertIt}), (\ref{eq:3}) are simply $\cN$=4 analogues of
this relation.

Restricted to the lowest order in the coupling constant ($g^2$ by the Feynman
rules in \cite{ESS,EKS2,EKS3}) this implies that the one-loop correction
to the original $n$-point correlator is equal to the integral over the
$(n+1)$-point function on the right hand side; we call this an ``operator
insertion''. We write expressions corresponding to Euclidean
Feynman rules, so for example the $i$ in front of the action is absorbed by Wick
rotation before the vertices are read off. 

In the ${\cal N} = 2$ formalism the left-handed odd variables $\q^\alpha$ carry $R$-charge $(+1)$ and
$\widehat W_{{\cal N} =2}$ is of charge (+2). Therefore the mixed correlator
${\cal G}_{n+1;1}(x_0;x_1,\dots,x_n)$ must be of order $\theta^4 +
O(\theta^5 \bar \theta)$.

The insertion relation (\ref{insertOneLoop}) is particularly simple to show
starting from the form of the Yang-Mills action given in (\ref{n2SYM}).
On the other hand, the $\theta = 0$ term of the $\widehat W_{{\cal N} = 2}$
multiplet
is one of the (complex) physical scalars of the ${\cal N} = 4$ SYM multiplet.
With the given field rescaling we find $\widehat W_{{\cal N} = 2} = g^{-1}
\, \phi(x) + O(\theta)$. In the following section it is implied that the field
redefinition by $g$ has been undone, so $\widehat W_{{\cal N} = 2, \ \mathrm{lin}} \rightarrow g \,
W_{{\cal N} = 2, \ \mathrm{lin}}$.

Last, in this appendix we have indicated the $U(1)$ charge assignments of the
harmonics and of the fields to be in accord with the literature. In the rest
of this work we simply write $q,\tilde q$ instead of $q^+, \tilde q^+$.

\section{ Reduction ${\cal N}=4 \, \rightarrow
{\cal N}=2$}
\label{sec:reduction-cal-n=4}
Real ${\cal N} = 4$ Minkowski space has the coordinates $x^{\alpha\dot \alpha},
\theta^{\alpha A},\bar \theta_A^{\dot \alpha}$ where $A \in \{1,\ldots,4\}$ and
$\alpha, \dot \alpha$ are the usual two-component indices. In order
to make touch with \cite{paulN4} we rather complexify, tacitly keeping
the notation $\theta, \bar \theta$ although the latter are
not complex conjugates of each other in the following.

The ${\cal N} = 4$ \emph{analytic superspace} of \cite{paulN4} has
additional coordinates $y_{a'}{}^a$ parametrising a coset of $GL(4)$:
\begin{equation}
\mathrm{Gr}(4,2) \, = \, \frac{GL(4,\mathcal{C})}{{\cal P}} \, = \, 
\left(
\begin{array}{rr}
\delta_b{}^a & 0 \\
y_{b'}{}^{a} & \delta_{b'}{}^{a'}
\end{array}
\right) \, = \, g_B{}^A
\end{equation}
where ${\cal P}$ is the (parabolic) subgroup of upper triangular matrices
with $2 \times 2$ blocks. We have split the indices as
\begin{equation}
A \, = \, ( a, a' ) \, , \quad a \in \{1,2\} \, , \quad  a' \in \{3,4\} \, .
\end{equation}
We can use these to project onto one half of the Grassmann coordinates:
\begin{equation}
\rho^{\alpha a} \, = \, \theta^{\alpha\, a} + \theta^{\alpha a'} \, y_{a'}{}^a \, , \qquad
\bar \rho_{a'}{}^{\dot \alpha} \, = \, y_{a'}{}^a \bar \theta_a^{\dot \alpha} +
\bar \theta_{a'}^{\dot \alpha}  \, .
\end{equation}

The field strength multiplet
\begin{equation}
W^{[AB]} \, = \, \phi^{[AB]}(x) + \theta^{\alpha[A} \psi(x)^{B]}_\alpha +
\theta^{[A}_{(\alpha} \theta^{B]}_{\beta)} F^{\alpha \beta} + O(\bar \theta)
\end{equation}
can also be projected by the ``harmonics''
\begin{equation}
W_{{\cal N} = 4}(x_\cA,\rho,\bar\rho,y) \, = \, \epsilon^{ab} g^a_A g^b_B W^{AB} \, .
\end{equation}
We see that the $\theta$ dependence is reduced to $\rho$ (similarly
only $\bar \rho$ remains).
The field strength multiplet thus lives on ``analytic superspace'' with
the coordinates $x^{\alpha \dot\alpha }_{\cA},\rho^{\alpha a},\bar\rho_{a'}{}^{\dot \alpha},
y_{a'}{}^a$. The change of basis $x \rightarrow x_\cA$ is analogous to the
${\cal N} = 2$ case. In particular, it involves $\bar \theta$ and so is
irrelevant in the present context.

The ${\cal N} = 2$ analytic superspace can be embedded into this larger space.
In order to reduce the field strength multiplet to ${\cal N} = 2$
pieces we need some of the $GL(4)$ raising operators, namely $D_{a'}^a$.
They act only on the $y$ variables according to
\begin{equation}
D^{a'}_a \, y_{b'}{}^b \, = \, \delta_{b'}^{a'} \delta_a^b \, .
\end{equation}
We define
\begin{eqnarray}
W_{{\cal N} = 4} & \rightarrow & q \nonumber \\
D^4_1 \, W_{{\cal N} = 4} & \rightarrow & \tilde q \label{eq:4}\\
D^3_1 \, W_{{\cal N} = 4} & \rightarrow & W_{{\cal N} = 2} \nonumber \, ,
\end{eqnarray}
so that for instance (recall $\cT = \tr(W_{{\cal N} = 4}^2)$)
\begin{eqnarray}
&& \frac{1}{8} \, (D^4_1|_2)^2 \, (D^4_1|_4)^2 \, D^4_1|_5 \, (D^3_1|_6)^2 \,
\la \cT_1 \cT_2 \cT_3 \cT_4 \cT_5 \cT_6 \ra \\ && \ \rightarrow \,
\la {O}_1 \, \tilde {O}_2 \, {O}_3 \, \tilde {O}_4 \, 
\widehat {O}_5 \, \tr(W_{{\cal N} =2}^2) \ra \, . \nonumber 
\end{eqnarray}
Further, we define $\tilde y_a^{a'} \, = \, \epsilon_{ab} \, \epsilon^{a'b'} \, y_{b'}^b$, so for lowering and raising of the two-component flavour indices the same
convention is used as in harmonic superspace. To add some detail:
\begin{equation}
\epsilon^{ab} \, \epsilon_{bc} \, = \, \delta^a_c \, , \qquad 
\epsilon_{a'b'} \, \epsilon^{b'c'} \, = \, \delta_{a'}^{c'} \, , \qquad
\epsilon_{12} \, = \, \epsilon_{34} \, = \, 1 \, .
\end{equation}
The symbol $y^2$ denotes the determinant of the matrix $y$,
\begin{equation}
y^2 \, = \, - \frac{1}{2} \, \tilde y^a_{a'} \, y_{a'}^a \, .
\end{equation}

On the level of the $y, \rho$ variables the reduction to ${\cal N}=2$ is
accomplished by
\begin{eqnarray}
& \quad y_3^2 \, \rightarrow \, \mathbf{y}
 \, , & y_3^1, y_4^1, y_4^2 \, \rightarrow \,
0 \\ & \theta^2, \theta^3 \, \rightarrow \, \theta^i \, , &  \theta^1,
\theta^4 \, \rightarrow \, 0 \nonumber
\end{eqnarray}
so in particular
\begin{equation}
\rho^a \, \rightarrow \, \delta^a_2 (\theta^2 + \theta^3 \, y_3^2) \, = \, 
\delta^a_2 \, \theta^i (1,\mathbf{y})_i \, = \, - \delta^a_2 \, \theta^+
\end{equation}
where we identified $(1,\mathbf{y})_i = u^+_i$. It follows
$(12) = u^{+i}|_1 u^+_i|_2 = \mathbf{y}_{12}$. Note that
\begin{equation}\label{eq:5}
y^2 \, \rightarrow \, 0 \, , \qquad D^4_1 \, y^2 \, \rightarrow \, \textbf{y}
\, , \qquad (D^4_1)^2 \, y^2 \, \rightarrow \, 0 
\end{equation}
as a consequence of the index contraction by the $\epsilon$ symbols.
Last, if we define $(\rho^2)_{(\alpha \beta)} = \epsilon_{ba} \rho_\alpha^a
\rho^b_\beta$ and similar for the ${\cal N} = 2$ variable $\theta^i_\alpha$:
\begin{equation}
(D^3_1)^2 \rho^4 \, = \, (D^3_1)^2 \frac{1}{12} \,
(\rho^2)^{(\alpha \beta)} (\rho^2)_{(\alpha \beta)}
\, = \, \frac{1}{6} \, (\theta^2)^{(\alpha \beta)} (\theta^2)_{(\alpha \beta)} \, = \,
2 \, \theta^4
\end{equation}

\section{
${\cal N}=4$ correlators from ${\cal N}=2$}
\label{sec:cal-n=4-correlators}

As mentioned previously perturbative computations of correlation functions are most easily performed in
$\cN$=2 harmonic superspace and are not possible directly in
$\cal N$=4 analytic superspace. But the amplitude/correlation function
duality naturally relates superamplitudes to correlation functions in
$\cal N$=4 analytic superspace. We here show how to
reconstruct the full $\cal N$=4 (bosonic part of the) correlator from various permutations of an
$\cal N$=2 correlator. 

We will first switch off all superspace
coordinates (corresponding to restricting ourselves to the MHV amplitudes). Of course in this paper we do not want to restrict ourselves to these cases, but they illustrate the procedure which we adapt in the main text to treat the various cases we are interested in. 
First we write down all allowed $y$ structures. In the $\cal N$=4 case
this means writing down all possible  products of $n$ $y_{ij}^2$ terms
such that each 
index occurs exactly twice (this is simply because the $R$-charge
of each operator, the energy momentum multiplet is two). Thus 
\begin{align}
  \label{eq:33}
  \vev{\cO\cO\dots \cO}= \sum_{\sigma \in S_n} y^2_{1 \sigma(1)} y^2_{2 \sigma(2)}
  \dots y_{n \sigma(n)}  f_\sigma(x)\ ,
\end{align}
where the sum is over all permutations of 1 to $n$ (in fact
``derangements'' - a derangement
being  a permutation in which no
element remains in its original position -  a permutation in which at
least one element $i$ remained fixed would
lead to $y^2_{ii}=0$). Note that different permutations may lead to
the same $y$-structure. The most obvious example of this is that
$\sigma$ and $\sigma^{-1}$ will always lead to the same
$y$-structure so that $f_\sigma=f_{\sigma^{-1}}$. However one can also
see that any cycle within a permutation may be replaced by its inverse
to give the same $y$-structure, so if $\sigma=\mu \nu$ then
$\sigma'=\mu^{-1}\nu$ also gives the same $y$-structure. Strictly
speaking we should consider equivalence classes of all such related
permutations but we won't worry too much about these details.

Such a correlator in $\cal N$=4 reduces to many different  $\cal N$=2
correlators.  {They correspond to the projections of the $\cN=4$ scalar operator $\cO$ onto $\cN=2$ operators made of hypermultiplet scalars, $\cO\ \to O, \, \tilde O, \,  \widehat O$ (see the definitions in \p{defop}), as well as projections made of the $\cN=2$ SYM scalar. } However we will show that the full $\cal N$=4 correlator can be
reconstructed entirely from specific types of $\cal N$=2 hypermultiplet 
correlator. More precisely for any term in the
full $\cal N$=4  correlator (\ref{eq:33}) (specified by a particular permutation
$\sigma$) we identify a (not necessarily unique)  $\cal N$=2
correlator which will give this term. The particular $\cal N$=2
correlator is determined as follows: write out the permutation
$\sigma$ as a product of disjoint cycles $\sigma=\sigma_1 \sigma_2
\dots \sigma_m$. Then construct an $\cal N$=2 correlator as follows:
put an operator $\cO$ at the point 
given by the first element of $\sigma_1$, an operator $\tilde O$ at the point given by the second position, and so on,
alternating between $O$ and $\tilde O$. If the cycle has
even length, then simply continue the procedure with the next
cycle. If however the 
cycle has odd length we must put an operator $\widehat O$ at the point
given by the last element of this odd cycle, then continue with the
next cycle. The coefficient of the $\cal N$=4 y-structure in question $f_\sigma$
can then be read off from the corresponding term in the $\cal N$=2
correlator. 

The procedure is best illustrated with an example. Say we wish to
determine the function $f_\sigma(x)$ in the $\cal N$=4, eight-point
correlator given by 
$y_{15}^2y_{14}^2y_{45}^2 y_{26}^4 y_{37}^2y_{38}^2y_{78}^2
f_\sigma(x)$. The permutation in question here can be given as
$\sigma=(154)(26)(378)$ (as mentioned above this is not unique, we
could have chosen (145) as the first cycle instead for example).
So according to the general procedure for determining an $\cal N$=2
correlator which will give this function, we put the operator  
$O$ at points 1,2,3, $\tilde O$ at points 5,6,7 and $\widehat O$ at
points 4,8 (corresponding to the last elements in the odd cycles). So
in other words
we consider the $\cal N$=2 correlator
\begin{equation}
  \label{eq:8}
  \vev{O O O \widehat O \tilde O \tilde O \tilde O \widehat O}
  = {\bf y}_{51}{\bf y}_{41}{\bf y}_{54} {\bf y}_{62}^2 {\bf
    y}_{37}{\bf y}_{38}{\bf y}_{87} f_{\sigma}(x)  + \dots
\end{equation}
where we have only displayed the relevant term in this $\cal N$=2
correlator which we are interested in. The important point is that the
$\cal N$=4 correlator 
reduces directly to this $\cal N$=2 correlator and the $\cal N$=4 y-structure
reduces directly to this $\cal N$=2 $y$-structure, thus the functions
$f_\sigma(x)$ are the same.\footnote{In order to ensure we don't get a minus sign error, it is important that we always write the $\cal N$=2 terms as  ${\bf y}_{ij} $ when $i$ is associated with $\tilde O$ and $j$ associated with $O$, rather than ${\bf y}_{ji}$.}

We conclude that we can reproduce any term in the $\cal N$=4
correlator by considering appropriate $\cal N$=2 correlators. We need
correlators with mostly $O$'s and $\tilde O$'s, but we also may
need a few correlators with $\widehat O$ operators. More precisely we need a $\widehat O$ for every
odd cycle in the permutation $\sigma$, the rest of the operators in the
correlator will be half $O$ and half $\tilde O$. 

So for example here we display all the types of $\cal N$=2 correlators needed
to reconstruct fully the (bosonic) $\cal N$=4 correlator for $n=3,4,5,6,7$ 
\begin{align}
  n=3 \qquad & \vev{O \tilde O \hat O} &&\rightarrow \qquad
\vev{\cO \cO \cO}\nonumber\\
 n=4 \qquad & \vev{O O \tilde O \tilde O}&&\rightarrow \qquad
\vev{\cO \cO \cO \cO}\nonumber\\ 
n=5 \qquad & \vev{O O \tilde O \tilde O \hat O }&&\rightarrow
\qquad \vev{\cO \cO \cO \cO \cO}\nonumber\\
n=6 \qquad & \vev{O O O \tilde O \tilde O \tilde O }+
\vev{O O \tilde O \tilde O \hat O \hat O } &&\rightarrow \qquad
\vev{\cO \cO \cO \cO \cO \cO}\nonumber\\
n=7 \qquad & \vev{O O O  \tilde O \tilde O \tilde O \hat O}
&&\rightarrow \qquad \vev{\cO \cO \cO \cO \cO \cO \cO}\ .
\end{align}
In particular we see that for $n=6 $ for the first time we need two different
types of correlator, the second type, with two $\hat O$'s is needed to
determine, $f_{\sigma}(x)$ whenever  $\sigma$ is a product of two three cycles,
which the first type of correlator will  miss.  

It is interesting to count the number of different terms in the correlators in
$\cal N$=4. The counting of
the number of independent $y$-structures is equivalent to counting
symmetric traceless $n\times n$ matrices $A$ with positive integer
entries, whose rows and columns add up to two. To see this imagine writing the
correlator as
$\vev{\cO \cO \dots \cO}= \sum_{A} \prod_{i,j=1}^n(y^2_{i j})^{A_{ij}/2}
  f_A(x)\ ,$ 
where the sum runs over the set of such matrices $A$.
The number of such structures for $n=2,3,4,5,6,7,8$ is $1, 1, 6, 22,
130, 822, 6202$ and one can find more details and references for the counting of such objects here \cite{oeis1}.

Finally in this paper we have been considering supercorrelators with
odd coordinates turned on which complicates the analysis. However  the
above techniques {\em can} be used to obtain the component $\la \cO(1) \dots \cO(n)
\cL  \dots \cL \ra$ with all $\rho_i=0\,, \ i=1\dots n$  from appropriate $\cN=2$ correlators, namely $\la
O(1) \dots O(n) \cL_{\cN=2} \dots \cL_{\cN=2} \ra$. Essentially the
Lagrangian components lift directly from $\cN=2$ to $\cN=4$ and the
rest lifts exactly as described above for the case with no Lagrangian
insertions.  It is this application which we make use of  a number of times in this paper.

\section{Relations between different superspace variables}
\label{sec:nairs-eta-as}

In this paper we make use of several different superspaces. The Grassmann odd
variables we use are Nair's $\eta$ and the momentum supertwistor variable $\chi$
(both familiar in the superamplitude context) and the analytic
superspace odd variable $\rho$ (useful for the correlation functions.)
Furthermore all of these variables can be defined in terms of the standard $\cN$=4
Minkoswski superspace variable $\theta$ which we have not made direct
use of here. Nevertheless it is clear that the variables are not
independent and we here give the relations between them, which are in
fact crucial for understanding the duality.   

{Firstly the variables $\chi$ are defined in terms of $\theta$ as~\cite{Mason:2009qx}
\begin{align}
  	\chi_i^A = \lambda_{i}^{\alpha} \theta_{i\alpha}^{ A} =  \lambda_{i}^{\alpha} \theta_{i+1\, \alpha}^{ A}\ .
\end{align}
Secondly the variables $\rho^{\alpha a}$ are simply harmonic projections of $\theta$ given explicitly  
as
\begin{align}
\rho_i^{\alpha a} = \q_i^{\alpha a} + \q_i^{\alpha a'} y_{i\,a'}{}^a\,.
\end{align}
These two relations together yield a direct  relation between $\chi$ and $\rho$ which we have made repeated use of~(\ref{chiToRho})
\begin{equation}
\chi_{i} \, = \, \la i| (\rho_i - \rho_{i \, i+1} \, y_{i \, i+1}^{-1} \, y_i) \, ,
\qquad
\chi_{i}' \, = \, \la i| \rho_{i \, i+1} \, y_{i \, i+1}^{-1} \, , \qquad
\la i| \, = \, \epsilon_{\alpha \beta} \, \lambda^\beta_i \, .
\end{equation}
}

Thirdly, a general formula relating the $\eta$ to the $\chi$ variables was given in
\cite{Mason:2009qx}:
\begin{equation}
\eta_{i}^A \, = \, \frac{\la i-1 \, i \ra \, \chi_{i-1}^A +
\la i \, i+1 \ra \, \chi_{i-1}^A + \la i+1 \, i-1 \ra \, \chi_{i}^A}
{\la i-1 \, i \ra \la i \, i+1 \ra} \ .
\end{equation}
We already have seen the relation between $\chi$ and
$\rho$~(\ref{chiToRho})
\begin{equation}
\chi_{i}' \, = \, \la i| \rho_{i \, i+1} \, y_{i \, i+1}^{-1} \, , \qquad
\chi_{i} \, = \, \la i| (\rho_i - \rho_{i \, i+1} \, y_{i \, i+1}^{-1}
\, y_i)\ .
\nonumber
\end{equation} 
Substituting this in we obtain the desired relation between $\eta$ and $\rho$
\begin{eqnarray}
\eta_i' & = & \frac{1}{\la i-1 \, i \ra} \la i-1 | \, \sigma_{i-1 \, i \, i+1}
- \frac{1}{\la i \, i+1 \ra} \la i+1 | \, \sigma_{i \, i+1 \, i +2} \, ,
\label{genEtaSol} \\
\eta_i & = & - \frac{1}{\la i-1 \, i \ra} \la i-1| \, \sigma_{i-1 \, i \, i+1} \,
y_i + \frac{1}{\la i \, i+1 \ra} \la i+1 | \, \sigma_{i \, i+1 \, i+2} \, y_{i+1}
\, , \nonumber
\end{eqnarray}
where
\begin{equation}
\sigma_{ijk} \, = \, \rho_{ij} \, y_{ij}^{-1} \, - \, \rho_{jk} \,
y_{jk}^{-1}\ .
\end{equation}

One can also see this directly from the relation between both variables and
the Minkowski superspace $\theta$:
Due to the matrix notation introduced in Appendix~B we can drop the Lorentz
and internal indices (writing $\theta'$ for $\theta^{a'}$) and rather
give the variables a point label. We thus have 
\begin{equation}
\rho_i \, = \, \theta_i \, + \, \theta_i' \, y_i \, . \label{defL}
\end{equation}
In the light-cone limit differences of $\theta^A$ can be expressed in terms
of $\eta^A$ and bosonic spinors:
\begin{equation}
\theta_{i,i+1} \, = \, |i\ra \, \eta_i \, , \qquad
\theta_{i,i+1}' \, = \, |i\ra \, \eta_i' \, \qquad x_{i,i+1} \, = \, |i\ra \,
[i|\ .
\end{equation}
>From the definition (\ref{defL}) we obtain
\begin{equation}
\theta_{12} \, = \, \rho_{12} - \theta_1' \, y_1 + \theta_2' \, y_2
\, = \,  \rho_{12} - \theta_1' \, y_{12} - \theta_{12}' \, y_2 \, .
\label{firstEtaEq}
\end{equation}
Our goal is to write $\eta_i,\eta_i'$ in terms of the analytic superspace
variables $\rho_i$ and $y_i$. Since we
want linearity we unfortunately have to keep $\theta_1'$ in the equation,
while the isolated $y_2$ does not look critical.

The R invariants first occur at five points, which thus seems to be a
natural and sufficiently non-trivial example. Proceeding like in
(\ref{firstEtaEq}) (which is repeated for completeness) we find the
system
\begin{eqnarray}
E_1 : \quad \theta_{12} & = & - \theta_{12}' \, y_2  + (- \theta_1' ) \, y_{12} +
\rho_{12}  \, , \nonumber \\
E_2 : \quad \theta_{23} & = & - \theta_{23}' \, y_3 + (\theta_{12}' - \theta_1')
\, y_{23} + \rho_{23} \, , \\
E_3 : \quad \theta_{34} & = & - \theta_{34}' \, y_4 + (\theta_{12}' + \theta_{23}'
- \theta_1') \, y_{34} + \rho_{34} \, , \nonumber \\
E_4 : \quad \theta_{45} & = & - \theta_{45}' \, y_5 + (\theta_{12}' + \theta_{23}'
+\theta_{34}' - \theta_1') \, y_{45} + \rho_{45} \, . \nonumber
\end{eqnarray}
The $\theta_{51} \, = \, \ldots$ condition is not independent, of course.
Putting in the light-cone variables the system becomes
\begin{eqnarray}
E_1 : \quad |1\ra \, \eta_1 & = & - |1\ra \, \eta_1' \, y_2  +
(- \theta_1' ) \, y_{12} +
\rho_{12}  \, , \nonumber \\
E_2 : \quad |2\ra \, \eta_2 & = & - |2\ra \, \eta_2' \, y_3 +
(|1\ra \, \eta_1' - \theta_1')
\, y_{23} + \rho_{23} \, , \\
E_3 : \quad |3\ra \, \eta_3 & = & - |3\ra \, \eta_3' \, y_4 +
(|1\ra \, \eta_1' + |2\ra \, \eta_2' - \theta_1') \, y_{34} + \rho_{34} \, ,
\nonumber \\
E_4 : \quad |4\ra \, \eta_4 & = & - |4\ra \, \eta_4' \, y_5 +
(|1\ra \, \eta_1' + |2\ra \, \eta_2' +  |3\ra \, \eta_3' - \theta_1') \, y_{45} +
\rho_{45} \, . \nonumber 
\end{eqnarray}
Every equation $E_i$ splits into two conditions because we can project with two
different bosonic spinors. We label
\begin{equation}
E_{ia} \, = \, \la i | \, E_i \, , \qquad E_{ib} \, = \, \la i+1 | \, E_i \, .
\end{equation}
These are eight equations whereas we try to solve for four $\eta_i$ and four
$\eta_i'$ and the two projections of $\theta_1'$, so a total of ten quantities.
What we can additionally invoke is the conservation condition on the
$\theta$'s. Splitting it into primed and un-primed halves and projecting
with $\la 4|, \, \la 5|$ we find the four conditions
\begin{eqnarray}
F_b : \quad \eta_4 & = & - \frac{\la 15 \ra}{\la 45 \ra} \eta_1 -
\frac{\la 25 \ra}{\la 45 \ra} \eta_2 - \frac{\la 35 \ra}{\la 45 \ra} \eta_3 \,
\nonumber \\
F_a : \quad \eta_4' & = & - \frac{\la 15 \ra}{\la 45 \ra} \eta_1' -
\frac{\la 25 \ra}{\la 45 \ra} \eta_2' - \frac{\la 35 \ra}{\la 45 \ra} \eta_3'
\, \\
\phantom{F_b : \quad} \eta_5 & = & \phantom{-} \frac{\la 14 \ra}{\la 45 \ra}
\eta_1 +
\frac{\la 24 \ra}{\la 45 \ra} \eta_2 + \frac{\la 34 \ra}{\la 45 \ra} \eta_3 \,
\nonumber \\
\phantom{F_b : \quad} \eta_5' & = & \phantom{-} \frac{\la 14 \ra}{\la 45 \ra}
\eta_1' +
\frac{\la 24 \ra}{\la 45 \ra} \eta_2' + \frac{\la 34 \ra}{\la 45 \ra} \eta_3'
\, \nonumber
\end{eqnarray}
The first two of these are the two missing conditions completing our system
to a total of ten equations. The other two then simply yield $\eta_5,\eta_5'$.
Solving the system is straightforward if cumbersome. We find
\begin{equation}
\la 1 | \theta_1' \, = \, \la 1 | \, \rho_{12} \, y_{12}^{-1} \, , \qquad 
\la 5 | \theta_1' \, = \, \la 5 | \, \rho_{51} \, y_{51}^{-1} \, .
\end{equation}
Finally, in terms of the $Q$ supersymmetric combination
\begin{equation}
\sigma_{512} \, = \, \rho_{51} \, y_{51}^{-1} \, - \, \rho_{12} \, y_{12}^{-1}
\end{equation}
the solution for $\eta,\eta'$ takes the simple form
\begin{eqnarray}
\eta_1' & = & \frac{1}{\la 51 \ra} \la 5 | \, \sigma_{512}
- \frac{1}{\la 12 \ra} \la 2 | \, \sigma_{123} \, , \label{etaSol} \\
\eta_1 & = & - \frac{1}{\la 51 \ra} \la 5 | \, \sigma_{512} \, y_1
+ \frac{1}{\la 12 \ra} \la 2 | \, \sigma_{123} \, y_2  \nonumber
\end{eqnarray}
and cyclic. Note that the un-primed $\theta$'s are unambiguously determined,
too, because $\theta_1 = \rho_1 - \theta_1' y_1$.

Our equations for $\eta$ in terms of $\rho$ carry over to the $n$-point case:
Splitting $\rho$ into $\theta,y$ makes (\ref{etaSol}) simplify to
$\eta^A = \eta^A$ for any number of points, so this is
a general solution. On the other hand, the number of odd degrees of freedom
always matches between the $\rho_i$ and the $\eta_i$, respectively.

\section{$\rho_i^4$ components of R invariants}
\label{sec:rho_i4-components-r}

A general formula for the R invariants~\cite{annecySuperspace} in terms of momentum supertwistor
variables was given in \cite{marksAndSparks}. The invariant is
characterise by 5 labels $r,s-1,s,t-1,t$.
\begin{align}
 R_{r,s-1,s,t-1,t} \label{NMH}
& = & \frac{\delta^4 \bigl( \Sigma_{r \, s-1 \, s \, t-1 \, t} \bigr)}{
\la s-1 \, s \, t-1 \, t \ra \la s \, t-1 \, t \, r \ra 
\la t-1 \, t \, r \, s-1 \ra \la t \, r \, s-1 \, s \ra
\la r \, s-1 \, s \, t-1 \ra}
\end{align}
with
\begin{equation}
\Sigma_{r \, s-1 \, s \, t-1 \, t} \, = \, \la s-1 \, s \, t-1 \, t \ra \,
\chi_r + (\text{cyclic}) \, . \label{5.8}
\end{equation}
The twistor four-bracket was defined in equation \p{detForm} in Section
\ref{numSec}. In the delta function in the numerator we cyclically shift
the five arguments $\la s-1 \, s \, t-1 \, t \ra \, \chi_r \, \rightarrow
\, \la s \, t-1 \, t \, r \ra \, \chi_{s-1}$ etc. yielding a total of five
terms.
The evaluation of this visually somewhat stunning formula at a
``Lagrangian point'' $\rho_j = 0 : j \neq i$ is surprisingly easy
because according to formula \p{chiToRho} only $\chi_{i-1}, \chi_i$
are non-vanishing. First, we observe
\begin{equation}\label{new}
\delta^{(4)}(\chi^A) \, = \, \chi^1 \, \chi^2 \, \chi^3 \, \chi^4 \, = \,
\frac{1}{4} \, \epsilon_{ab} \chi^a \chi^b \; \epsilon_{a'b'}
\chi^{a'} \chi^{b'} \, = \, \frac{1}{4} \, (\chi)^2 (\chi')^2 \, .
\end{equation}
Let us focus on the case $R_{12345}$ at $\rho_5^4$.
>From \p{chiToRho} we have
\begin{equation}
\chi_4' \, = \, - \la 4 | \rho_5 \, y_{45}^{-1} \, , \qquad
\chi_5' \, = \, \la 5 | \rho_5 \, y_{51}^{-1}
\end{equation}
and
\begin{equation}
\chi_4 \, = \, - \chi_4' \, y_4 \, , \qquad \chi_5 \, = \, - \chi_5' \, y_1 \, . 
\end{equation}
Then
\begin{equation}
\delta^{(4)} \bigl( X \, \chi_{4}^A + Y \, \chi_{5}^A \bigr) \, = \,
\frac{1}{4} \, \bigl( X \, \chi_4 + Y \, \chi_5 \bigr)^2
\bigl( X \, \chi_4' + Y \, \chi_5' \bigr)^2 \, .
\end{equation}
The second square on the right-hand side is equivalent to $\delta(X \,
\chi_4' + Y \, \chi_5')$ so that we can rewrite the first square as
$(X \chi_4' y_{41})^2$. This in turn sends $\chi_4'$ to zero in the second term,
whence
\begin{eqnarray}
&& \delta^4 \bigl( X \, \chi_{4}^A + Y \, \chi_{5}^A \bigr) \, = \, 
\frac{1}{4} \, (X \, \chi_4' \, y_{41})^2 \, (Y \, \chi_5')^2 \, = \, 
\frac{1}{4} \, X^2 \, Y^2 \, y_{14}^2 \, (\chi_4')^2 (\chi_5')^2
\label{doubleDelta} \\
&& \quad = \frac{1}{4} \, X^2 \, Y^2 \frac{y_{14}^2}{y_{45}^2 \, y_{51}^2} \,
(\la 4 | \rho_5)^2 (\la 5| \rho_5)^2 \, = \, X^2 \, Y^2 \,
\la 4 5 \ra^2 \, \frac{y_{14}^2}{y_{45}^2 \, y_{51}^2} \, \rho_5^4 \nonumber
\end{eqnarray}
where we have put $\rho^4 \, = \, \frac{1}{12} \, (\rho^2)^{(\alpha \beta)}
(\rho^2)_{(\alpha \beta)}$ as before.

The five-point case is somewhat degenerate because all five $\chi_i$ occur
in a cyclic fashion in the numerator, whereby the labelling is arbitrary.
To be definite let $r=1$. We find immediately
\begin{equation}
R_{12345}\vert_{\rho_5^4} \, = \, \frac{\la 45 \ra^2 \la 5123 \ra \la 1234 \ra}
{\la 2345 \ra \la 3451 \ra \la 4512 \ra} \, \frac{y_{14}^2}{y_{45}^2 \,
y_{51}^2} \, \rho_5^4 \, = \, \frac{x^2_{13} x^2_{24}}{x^2_{14} x^2_{25} x^2_{35}}
\, \frac{y_{14}^2}{y_{45}^2 \, y_{51}^2} \, \rho_5^4 \label{resR135}
\end{equation}
where \p{detToProp} was used to translate to $x$ space.

At six points or above, Lagrangian components of the R invariants may
vanish: In some constellations only one or no non-vanishing $\chi$ will
be amongst the five terms in the argument of the delta function. As
in the main text, at six points we label the $R_{r,s-1,s,t-1,t}$ by the
missing point, so e.g. $R_1 = R_{23456}$. For a start, we wish
to rewrite $R_{1}|_{\rho_6^4}$ above in terms of $x$ space variables.
Using \p{doubleDelta} we can immediately rewrite as follows:
\begin{align}
  \label{eq:54}
  R_{1}= {x_{35}^2 \over x_{46}^2 x_{36}^2} {\vev{6234}
    \vev{45}\over \vev{4562}\vev{34}} {y_{15}^2 \over y_{16}^2
    y_{56}^2} \rho_6^4= {x_{35}^2 \over x_{46}^2 x_{36}^2}  {y_{15}^2 \over y_{16}^2
    y_{56}^2} \rho_6^4 \times I
\end{align}
To do this note that $I \, x_{15}^2/x_{14}^2$ is conformally invariant:
\begin{align}
  \label{eq:55}
  {x_{15}^2\over x_{14}^2}  I  = {\vev{6234} \vev{6145} \over \vev{4562}
  \vev{6134}} = {z_{51}z_{23} \over z_{13} z_{25}} \, .
\end{align}
Here we have substituted the momentum twistor four-brackets with the
variables $z_i$ as discussed 
previously in Section~\ref{sec:transl-r-invar-1} in the main text.
The one-dimensional cross-ratio can be rewritten in terms of the standard
six-point cross-ratios:
\begin{align}
  \label{eq:56}
  u_1={x_{31}^2 x_{46}^2 \over x_{36}^2 x_{4 1}^2}={z_{12}z_{45} \over
    z_{14}z_{25}} \, , \qquad   u_2={x_{15}^2 x_{24}^2 \over x_{14}^2 x_{2
        5}^2}={z_{23}z_{56} 
      \over z_{25}z_{36}} \, , \qquad   u_3={x_{26}^2 x_{35}^2 \over
        x_{25}^2 x_{3 6}^2}={z_{34}z_{61} \over z_{36}z_{41}}
\end{align}
to give
\begin{align}
  \label{eq:57}
  {x_{15}^2\over x_{14}^2} \, I = {1-u_1+u_2-u_3+ \sqrt{\Delta}
 \over 2 (1-u_3)} \, , \qquad \Delta=(1-u_1-u_2-u_3)^2 - 4 \, u_1 u_2 u_3 \, .
\end{align}
To obtain this expression, the simplest way is to use one-dimensional
conformal invariance: Set, say $z_1=0, z_2=\infty, z_3=1$ so
that (\ref{eq:56}) determines $u_1,u_2$ and $u_3$ in terms of
$z_4,z_5, z_6$ and then take the same limit in (\ref{eq:55}) replacing
the $z$ with the $u$. Conversely, in the $z$ variables $\Delta$ becomes
a perfect square so that it is easy to verify the last formula.
Collecting terms we get
\begin{align}
  \label{eq:60}
R_1|_{\rho_6^4}= {x_{35}^2x_{14}^2 \over x_{46}^2 x_{36}^2
x_{15}^2} \Big({1-u_1+u_2-u_3 + \sqrt{\Delta} \over 2 (1-u_3)}  \Big)
{y_{15}^2 \over y_{16}^2 y_{56}^2} \rho_6^4 \ .
\end{align}
In a similar fashion we obtain
\begin{align}
  R_3|_{\rho_6^4}&= {x_{35}^2x_{14}^2 \over x_{46}^2 x_{36}^2x_{15}^2}
  \Big({1+u_1-u_2-u_3 - \sqrt{\Delta}  \over 2 u_3(1-u_3)}  \Big)  {y_{15}^2
\over y_{16}^2 y_{56}^2} \rho_6^4 \, , \nonumber \\
R_5|_{\rho_6^4}&=0 \, , \nonumber \\
 R_2|_{\rho_6^4}&= {x_{35}^2x_{14}^2 \over x_{46}^2 x_{36}^2x_{15}^2}
  \Big({1-u_1-u_2+u_3 - \sqrt{\Delta}  \over 2 u_3(1-u_1)}  \Big)  {y_{15}^2
\over y_{16}^2 y_{56}^2} \rho_6^4 \, , \label{citePaul}\\
 R_4|_{\rho_6^4}&= {x_{35}^2x_{14}^2 \over x_{46}^2 x_{36}^2x_{15}^2}
\Big({u_1(1-u_1+u_2-u_3 + \sqrt{\Delta})  \over 2 u_3(1-u_1)}  \Big) 
{y_{15}^2 \over y_{16}^2 y_{56}^2} \rho_6^4 \, , \nonumber \\
R_6|_{\rho_6^4}&=0 \, . \nonumber
\end{align}

\end{document}